\documentclass[aps,prd,eqsecnum,reprint,amsmath,amssymb,amsfonts,nofootinbib,superscriptaddress]{revtex4-2}

\usepackage{amsfonts,amssymb,amsmath}
\usepackage{xr}
\usepackage{graphicx, hyperref, tabularx}
\usepackage{microtype}

\newcommand{\dual}{\,{}^*\!}
\usepackage{bbold}
\renewcommand{\t}{{\mathbb{t}}}  %for tangential vector
\newcommand{\one}{{\mathbb{1}}}  %identity matrix
\newcommand{\Bbf}{\mbox{$\mathbf{B}$}}
\newcommand{\Rbf}{\mbox{$\mathbf{R}$}}

\newcommand{\xbf}{\mbox{$\mathbf{x}$}}
\newcommand{\lbf}{\mbox{$\mathbf{l}$}}
\newcommand{\Lbf}{\mbox{$\mathbf{L}$}}

\usepackage{xcolor}
\definecolor{mygray}{gray}{0.9}

\newcommand{\lrr}{Liv. Rev. Relativity}
\newcommand{\mnras}{Mon. Not. R. Astron. Soc.}
\newcommand{\cqg}{Class. Quantum Grav.}
\newcommand{\apjs}{Astrophys. J. Suppl.}

\begin{document}

\title{Characteristic Decomposition for Relativistic Numerical Simulations:\\
II. Magnetohydrodynamics}

\author{Saul A. Teukolsky}
\email{saul@astro.cornell.edu}
\affiliation{Cornell Center for Astrophysics and Planetary Science,
  Cornell University, Ithaca, NY 14853, USA}
\affiliation{Theoretical Astrophysics 350-17,
  California Institute of Technology, Pasadena, CA 91125, USA}
\date{\today}

\begin{abstract}
The characteristic decomposition for GRMHD in the comoving frame
of the fluid has been known for a long time. However, it has
not been known in the coordinate frame of the simulation and in
terms of the conserved variables evolved in typical numerical
simulations. This paper applies the method of quasi-invertible
transformations developed in Paper I to derive this decomposition.
Among other benefits, this will allow us to use the most accurate
known computational methods, such as full-wave Riemann solvers.
The results turn out to be simpler than expected based on earlier
attempts.
\end{abstract}

\maketitle

\section{Introduction}

Relativistic flows with strong magnetic and gravitational fields
are associated with extremely energetic astrophysical phenomena, such
as pulsar winds, anomalous X-ray
pulsars, soft gamma-ray repeaters, gamma-ray bursts, and
jets in active galactic nuclei. The merger of two neutron stars
or a neutron star and a black hole is another arena where we need
to be able to deal with this kind of physics.

Much of the time, the magnetic field can be treated in
the ideal magnetohydrodynamic limit (MHD), where the fluid is treated
as a perfect conductor.
General relativity (GR)
describes the gravitational field and modifies the equations
of MHD from their non-relativistic form. GRMHD is used both
to model LIGO mergers involving neutron stars and to match
observations with the Event Horizon
Telescope.

Numerical simulations with GRMHD are much more challenging than
purely hydrodynamic simulations in GR. There are more equations and
they are more complicated. In addition, there are physical instabilities
that can require very high resolution to study accurately. And the
conservation law that the divergence of the $\Bbf$-field
must vanish
can be difficult for the evolution equations
to satisfy numerically. The numerical methods
currently used are variants of high-resolution shock-capturing methods
first developed for hydrodynamics.

There is a serious remaining 
theoretical gap in the implementation of all GRMHD codes, both with
finite-difference and discontinuous Galerkin techniques.
The gap is the lack of a complete characteristic decomposition for the system
of equations. 
In numerical relativity, the evolution of fluids, magnetic fields,
and the metric itself
is done with hyperbolic evolution equations. Hyperbolicity
means that at every point the field admits a
decomposition into independent waves, each with its own
wave speed.  Formally, the
decomposition is carried out by solving an eigenvalue problem.
The most accurate numerical algorithms typically
make use in some way of this characteristic decomposition.
These uses include accurate imposition of boundary conditions
and treatment of shock waves.
Moreover, the most
robust and accurate prescriptions for numerical fluxes
across subdomain boundaries are based on characteristic
decomposition.

The characteristic decomposition of the MHD equations in general relativity
was first studied by Bruhat~\cite{bruhat1966} and then by Anile and
Pennisi~\cite{anile_pennisi1987}. The Anile and Pennisi work
uses essentially the same form of the equations as Bruhat,
except for changing one of the independent thermodynamic
variables from enthalpy to pressure. This work
was reproduced in Anile's book~\cite{anile1989}, from which the results have 
been widely used in the literature.
We note that,
although Bruhat was the first to
completely analyze the MHD characteristic speeds in full general
relativity directly from the characteristic matrix,
her work has never
been cited in this connection by subsequent authors,
with the exception of Ref.~\cite{vanputten1991}.

A key result of the Anile and Pennisi work is that there is a formulation
of the MHD equations that is strongly
hyperbolic, a necessary condition for having a well-posed initial-value
problem and thus being suitable for numerical computations.
However, the equations studied in this early work are
not in the form typically used in current numerical work.

Virtually all numerical codes today use equations in conservation
form: A set of evolved variables is chosen so that their spatial
derivatives appear only in the divergences of fluxes. Equations
in conservation form allow a wide variety of robust and
accurate treatments of shocks and other
discontinuities in the flow.
As shown in Ref.~\cite{schoepe2018},
the equations in conservation form differ from those of Bruhat
and Anile by the addition
of multiples of the divergence constraint $\nabla\cdot\Bbf=0$.
This does not change the solutions, but does change the
principal part of the equations (the highest derivative terms of the
evolved variables). Changing the principal part changes the
eigenvalues and eigenvectors of the characteristic matrix,
and hence can affect the hyperbolicity of the system.
In fact, Ref.~\cite{schoepe2018} shows that the usual conservative
system, the so-called Valencia formulation
(see, e.g.~the review in~\cite{font2008}), is only weakly hyperbolic
and thus does not have a well-posed initial-value problem.

One of the challenges of MHD simulations, both relativistic and in the
Newtonian limit, is to guarantee that the divergence constraint remains
satisfied during the evolution. Various computational techniques
have been developed to ensure this, one of which is divergence
cleaning~\cite{moesta2014,liebling2010,penner2011}.
In this method, we introduce a scalar field with an evolution
equation that damps the divergence of the
magnetic field and also advects it off the
grid.
Interestingly, Ref.~\cite{hilditch2019} has shown that the
divergence cleaning formulation of the Valencia scheme \emph{is}
strongly hyperbolic and thus has a well-posed initial-value problem,
making it attractive for numerical simulations.

In Paper I~\cite{teukolsky2025a}, we introduced a new method for
finding characteristic decompositions of fluid flows in relativity,
which we describe further below. The goal of the current paper is to apply
this method to GRMHD.
Besides the work of Anile~\cite{anile1989}, the
previous works on the characteristic decomposition problem for
GRMHD that are most related to this paper are those of
Anton et al.~\cite{anton2010}, Ibanez et al.~\cite{ibanez2015},
Schoepe et al.~\cite{schoepe2018},
and Hilditch and Schoepe~\cite{hilditch2019}.
We will discuss the connections at appropriate points in the text.
A detailed comparison with the work of Anton et al.~\cite{anton2010}
is given in Appendix \ref{app:anton} amd with that of
Schoepe et al.~\cite{schoepe2018} in Appendix \ref{app:schoepe}.
For other references to earlier work, see for example the introduction to
Ref.~\cite{anton2010}.

As discussed in more detail in Paper I~\cite{teukolsky2025a},
in the 3+1 decomposition
we introduce a coordinate frame and consider the evolution
with respect to the $t=\text{constant}$ slices of this frame.
The tangent vectors $t^a=(\partial/\partial t)^a$ to the $t$-coordinate
lines are related to the normal $n^a$ to the slices by
\begin{equation}
t^a=\alpha n^a+\beta^a.
\end{equation}
Here $\alpha$ is the lapse function and $\beta^a$ is the shift
vector.\footnote{In this paper, Roman letters $a,b,\ldots$
from the beginning of the alphabet label 4-d abstract spacetime indices.
Roman letters $i,j,\ldots$ from the middle of the alphabet
label 3-d spatial indices in some particular coordinate system.}

A special case is to choose the comoving or Lagrangian frame,
where the shift vector is chosen so that the spatial coordinates
are comoving with the fluid.
In this case $t^a$ is proportional to $u^a$, the 4-velocity of the fluid.
As noted in Paper I and by many previous authors,
the characteristic decomposition for hydrodynamics or MHD is most
easily determined in this frame, since
it is the natural frame associated with the fluid flow.
Ultimately, however, we will want the characteristic decomposition
in the 3+1 coordinate frame
and for the conserved variables in that frame.
We noted in Paper I that
we can first find the decomposition in the Eulerian frame, which
is the frame of observers at rest in the $t=\text{constant}$ slices
and whose 4-velocity is simply $n^a$.
Then Eq.~(\ref{I-eq:eigs_related}) of Paper I
shows that the eigenvalues in the 3+1 coordinate frame are simply related
to those in the Eulerian frame, while the eigenvectors are the same.
So it is sufficient to find the decomposition in the Eulerian frame.

In Paper I, we introduced a method of doing this entire
procedure for hydrodynamics
by first finding the
decomposition for a set of primitive variables in the comoving frame and
then transforming the eigensystem to
the conserved Eulerian variables.
The key difference between this
algorithm and previous transformation techniques is the use of
a \emph{quasi-invertible transformation}. The main technical innovation
in the current paper is the extension of
the quasi-invertible transformation procedure from the relatively simple
case of the fluid velocity to the more challenging case of the
magnetic field. We will first
apply the technique to the non-conservative
Anile formulation, even though this formulation is not
used in numerical work. This will make contact with previous literature
and is also useful because many of the results are identical for the
conservative divergence cleaning formulation, which we study afterward.

This paper is organized as follows: In \S\ref{sec:noncons} we
derive a system of equations equivalent to those of
Anile\cite{anile1989} for the nonnonservative formulation of
GRMHD. We then carry out the comoving decomposition
in \S\ref{sec:comovingdecomp} in enough detail to make the
subsequent developments clear. \S\ref{sec:quasiinvert} contains
the main technical accomplishment of the paper, the quasi-invertible
transformation from the comoving variables $\{u^a,b^a\}$ to
the Eulerian variables $\{v^a,B^a\}$ and vice versa. This transformation
is then applied to the comoving decomposition to get the
decomposition in the Eulerian frame. \S\ref{sec:consvariables} introduces
the conserved variables. The Jaocobian transformation matrix from
primitive to conserved variables is given in \S\ref{sec:transmatright}.
The inverse transformation is given in \S\ref{sec:transmatleft} in
a form much simpler than in previous work. These
transformations are then applied to the Eulerian Anile eigenvectors
to express them in terms of conserved variables. These
are not the eigenvectors for the conservative evolution equations,
but make contact with earlier work and will be useful in
presenting the correct conservative decomposition.
We present this decomposition for the conservative
formulation with divergence cleaning in \S\ref{sec:divclean}.

\section{Nonconservative Formulation of Magnetohydrodynamics}
\label{sec:noncons}

In the MHD approximation, the electromagnetic field is completely
determined by the comoving magnetic field
\begin{equation}
b^a=-\dual F^{ab}u_b,\qquad \dual F^{ab}=u^b b^a-u^a b^b.
\end{equation}
Here $\dual F^{ab}$ is the dual of the electromagnetic field tensor:
\begin{equation}
\dual F^{ab}=\tfrac{1}{2}\epsilon^{abcd} F_{cd}.
\end{equation}
Note that all electromagnetic
field quantities in this paper are renormalized by a factor $1/\sqrt{4\pi}$,
the standard convention in numerical work. Also, note that the
Levi-Civita tensor $\epsilon_{abcd}$ is defined with the
convention~\cite{GravitationMTW}
that in flat spacetime $\epsilon_{0123}=+1$.
The Maxwell equation $\nabla_a \dual F^{ab}=0$ becomes\footnote{We insert
the minus sign in front of this equation so that the corresponding
term on the diagonal of the matrix $A^a$ in Eq.\ \eqref{eq:amatrix} is
positive, the same as the sign of the other diagonal terms.
This simplifies the orthogonality relation for the eigenvectors
(cf.\ Eq.\ (\ref{I-eq:orthog}) of Paper I.)}
\begin{equation}
-\nabla_a(u^b b^a-u^a b^b)=0.
\label{eq:maxwell}
\end{equation}
The projection of this equation along $u_b$ gives the divergence
constraint, which can be written in several equivalent forms:
\begin{subequations}
\begin{align}
\label{eq:cons1}
\nabla_a b^a+u^a u^b\nabla_b b_a&=0,\\
\label{eq:cons2}
\nabla_a b^a-b^a u^b\nabla_b u_a &=0,\\
\label{eq:cons3}
h^{ab}\nabla_b b_a&=0.
\end{align}
\end{subequations}
Here $h^{ab}$ is the projection tensor that projects orthogonal
to $u^a$:
\begin{equation}
h^{ab}=g^{ab}+u^a u^b.
\end{equation}
Equation \eqref{eq:cons2} follows from Eq.\ \eqref{eq:cons1} using $u^ab_a=0$.
Equation \eqref{eq:cons3} follows from Eq.\ \eqref{eq:cons1} using
$\nabla_a b^a=(h^{ab}-u^a u^b)\nabla_a b_b$. As we will see below,
the divergence constraint plays an important role in distinguishing
different formulations of the MHD equations.

We will start out using $u^a$ and $b^a$ as two of the primitive variables.
In addition, we need two variables describing the
thermodynamic state for a perfect fluid.
Of course, at some stage we have to enforce the constraints
\begin{equation}
u^a u_a=-1,\qquad u^a b_a=0,
\label{eq:constraints}
\end{equation}
reducing the number of independent variables from 10 to 8.

It does not make much difference which two thermodynamic variables
we choose as primitive variables; it is easy to transform from
one set to any other. We will choose the pair
$(p,\epsilon)$, where $p$ is the pressure and $\epsilon$ the
specific internal energy density. This choice simplifies the
algebra slightly.
Later we will replace $p$ by $\rho$, the rest-mass density,
since $(\rho,\epsilon)$ are more convenient
in astrophysical codes.

Associated with the fluid are also
the total energy density $e$,
the specific entropy $s$, 
and the specific enthalpy $h$.
These quantities are related by the following
equations:
\begin{align}
\label{eq:e}
e&=\rho(1+\epsilon),\\
h&=1+\epsilon+p/\rho,\\
\label{eq:ds}
T\,ds&=d\epsilon+p\,d(1/\rho)\quad\text{(First Law)},\\
p&=p(\rho,\epsilon)\qquad\text{(Equation of state)},\\
dp&=
\left.\frac{\partial p}{\partial \rho}\right|_\epsilon d\rho +
\left.\frac{\partial p}{\partial \epsilon}\right|_\rho d\epsilon 
\equiv \chi d\rho + \kappa d\epsilon.
\label{eq:chi}
\end{align}
Here $T$ is the temperature.
The quantities $\chi$ and
$\kappa$ are related
to the speed of sound $c_s$ by
\begin{equation}
c_s^2=\frac{1}{h}\left(\chi+\frac{p}{\rho^2}\kappa\right)
\label{eq:cs}
\end{equation}
(cf.\ e.g., Appendix \ref{I-app:a} of Paper 1).

We now present a formulation of the MHD equations equivalent to
that of Anile~\cite{anile1989}, but in slightly simpler form. In
particular, there are only 8 dynamical equations since
the constraints  $u^a u_a=-1$ and
$u^a b_a=0$ are explicitly enforced. The formulation here is
equivalent to that in Ref.~\cite{schoepe2018}.

A basic equation for fluid flow is the equation of mass conservation,
\begin{equation}
u^a\nabla_a \rho+\rho \nabla_a u^a=0.
\label{eq:rhocons}
\end{equation}

The next fundamental equation is the energy equation. This can
be derived from the projection of the conservation equation $\nabla_a T^{ab}$
along $u_b$, where $T^{ab}$ is the total stress-energy tensor. It is
simpler to proceed as follows: Since the fluid is
taken to be perfect, the flow is isentropic (except at shocks).
Accordingly, we set $ds=0$ in Eq.\ \eqref{eq:ds} and get
\begin{equation}
\frac{d\epsilon}{d\tau}=\frac{p}{\rho^2}\frac{d\rho}{d\tau}.
\label{eq:epsrho}
\end{equation}
Here $\tau$ is the proper time along the fluid worldlines. Rewrite
$d/d\tau=u^a \nabla_a$ and use Eq.\ \eqref{eq:rhocons}
so that the energy equation becomes
\begin{equation}
u^a\nabla_a \epsilon+\frac{p}{\rho}\nabla_a u^a=0.
\label{eq:epsilon}
\end{equation}

To use $p$ as a primitive variable, we need its evolution equation.
Equation \eqref{eq:chi} gives
\begin{align}
\frac{dp}{d\tau}&=
\chi \frac{d\rho}{d\tau} + \kappa \frac{d\epsilon}{d\tau}\notag\\
&=\left(\chi+\kappa\frac{p}{\rho^2}\right)\frac{d\rho}{d\tau}\notag\\
&=h c_s^2 \frac{d\rho}{d\tau},
\end{align}
where we have used Eqs.\eqref{eq:epsrho} and \eqref{eq:cs}. So finally
\begin{equation}
u^a\nabla_a p + \rho h c_s^2\,\nabla_a u^a=0.
\label{eq:pressure}
\end{equation}

The total stress-energy tensor for the matter plus magnetic field is
\begin{equation}
T^{ab}=\rho h^* u^a u^b+p^* g^{ab}-b^ab^b,
\end{equation}
where
\begin{equation}
\rho h^*=\rho h +b^2, \qquad p^*=p+\tfrac{1}{2}b^2.
\end{equation}
The Euler equation follows from projecting the conservation equation
$\nabla_b T^{ab}=0$ orthogonal to $u^a$ using
$h^{ab}$.
This gives
\begin{multline}
\rho h^* h^d{}_b u^a\nabla_a u^b + h^{da}\nabla_a p\\
 + h^d{}_b b^a(\nabla^b b_a-\nabla_a b^b)-b^d\nabla_a b^a=0.
\label{eq:euler1}
\end{multline}
The term proportional to $\nabla_a b^a$ in this equation turns out to be
problematic in determining the characteristic decomposition below
because the derivative contains pieces both along and perpendicular
to $u^a$. Accordingly, the characteristic matrix is not completely
orthogonal to $u^a$.
We eliminate this problematic term by forming
the combination
\begin{equation}
b^a\nabla_b T_a{}^b+ \rho h^* \text{(l.h.s.\ of Eq.\ \ref{eq:cons2})}=0.
\label{eq:combo}
\end{equation}
This equation simplifies to 
\begin{equation}
\nabla_ab^a=-\frac{1}{\rho h}b^a \nabla_a p.
\label{eq:divb}
\end{equation}
Note that
the $u^a$ component of the
derivative on the right-hand side of Eq.\ \eqref{eq:divb}
drops out because it is projected along $b^a$.
Substituting Eq.\ \eqref{eq:divb} in Eq.\ \eqref{eq:euler1} gives the final
form
\begin{multline}
\rho h^* h^d{}_b u^a\nabla_a u^b + h^{da}\nabla_a p\\
 + b^a\left[h^d{}_b
(\nabla^b b_a-\nabla_a b^b) +\frac{b^d}{\rho h}\nabla_a p\right] =0.
\label{eq:euler2}
\end{multline}
This equation is equivalent to Eq.\ (2.75) of Ref.\ \cite{anile1989} when
that equation is projected with $h_{ab}$; the component of that equation
along $u_a$ is zero.

We used  the component of the Maxwell equation \eqref{eq:maxwell} along
$u_b$ in Eq.\ \eqref{eq:combo}, so we only need the projection with $h_{cb}$.
This gives
\begin{equation}
-h_{cb}(b^a\nabla_a u^b-u^a\nabla_a b^b)+b_c \nabla_a u^a=0.
\label{eq:maxwell2}
\end{equation}
This equation is equivalent to Eq.\ (2.76) of Ref.\ \cite{anile1989} when
that equation is projected with $h_{ab}$ and Eq.\ \eqref{eq:pressure} is used.
Again, note that the component of Eq.\ (2.76) in
Ref.\ \cite{anile1989} along $u_a$ is zero.

To enforce the condition $u^a u_a=-1$, in the above equations
we use the identity
\begin{equation}
\nabla_a u^a=h^a{}_b \nabla_a u^b.
\label{eq:div}
\end{equation}
Then Eqs.\ \eqref{eq:euler2}, \eqref{eq:maxwell2}, \eqref{eq:pressure} and
\eqref{eq:epsilon}
can be written as a quasi-linear system (Eq.~\ref{I-eq:system} of Paper I)
for the variables $(u^a,b^a,p,\epsilon)$,
where the matrices $A^a$ are
\begin{widetext}
\begin{equation}
A^a=
\begin{bmatrix}
u^a h_c{}^b & (b^b h_c{}^a -b^a h_c{}^b)/(\rho h^*) &
 h_c{}^a/(\rho h^*)+b_c b^a/(\rho^2 h h^*) & 0\\[3pt]
-b^a h_c{}^b +b_c h^{ba} & h_c{}^b u^a & 0 & 0\\[3pt]
\rho h c_s^2 h^{ba} & 0    & u^a    & 0\\[3pt]
(p/\rho)h^{ba} & 0 & 0   & u^a
\end{bmatrix}.
\label{eq:amatrix}
\end{equation}
\end{widetext}

\subsection{Comoving Decomposition}
\label{sec:comovingdecomp}

\subsubsection{Characteristic speeds}
\label{sec:charac}
As explained in Section \ref{I-sec:decomp} of Paper I,
to find the characteristic speeds in the comoving frame,
we introduce a vector
\begin{equation}
q_a=y_u u_a+\zeta_a,
\label{eq:qu}
\end{equation}
where $y_u$ is the comoving eigenvalue (characteristic speed) and $\zeta_a$ is
a unit vector orthogonal to $u_a$ defining the direction of the
decomposition.
We then set to zero the determinant
of the characteristic matrix $A^a q_a$.
The determinant is evaluated in Appendix \ref{app:a} and we find up to
a constant factor the well-known result
\begin{multline}
\det(A^a q_a)=a^2(B^2-\rho h^* a^2)\\
  \times\{a^2G b^2-\rho h a^4+c_s^2[\rho h a^2(a^2+G)- B^2 G]\},
\label{eq:detfinal}
\end{multline}
where we use the common notation
\begin{equation}
a=q^a u_a,\quad B=q^a b_a,\quad G=q^a q_a.
\label{eq:defns}
\end{equation}

From Eq.\ \eqref{eq:detfinal}, we see that there are three classes
of characteristic speeds:
\begin{itemize}
\item
There is a doubly degenerate eigenvalue given by $a=u^a q_a=y_u=0$.
\item
Two \emph{Alfv\'en waves} have speeds given by $B=\pm\sqrt{\rho h^*}a$.
With our conventions, the choice $B=-\sqrt{\rho h^*}a$ leads to
$y_u=+b^a\zeta_a/\sqrt{\rho h^*}$, which we will take to be the ``first''
Alfv\'en eigenvector.
\item
Four \emph{magnetosonic waves} have speeds given by the roots of\\
\newlength{\eqnlength}
\setlength{\eqnlength}{\linewidth}
\addtolength{\eqnlength}{-\leftmargin} %to avoid eqn being too long
\parbox[c]{\eqnlength}{%
\setlength{\multlinegap}{0pt}%
\setlength{\abovedisplayskip}{0pt}%
\setlength{\belowdisplayskip}{0pt}%
\begin{multline}
N\equiv a^2G b^2-\rho h a^4+c_s^2[\rho h a^2(a^2+G)- B^2 G]=0,\\
\label{eq:N}
\end{multline}}
a quartic in $q^a$.
\end{itemize}

\subsubsection{Right Eigenvectors in the Comoving Frame}
\label{sec:comoving}

The right eigenvectors follow from solving
$(A^a q_a)X=0$. Take $X$ to be of the form
\begin{equation}
X=\begin{bmatrix}
X_b\\[3pt]
Y_b\\[3pt]
X_7\\[3pt]
X_8
\end{bmatrix}.
\label{eq:Xvec}
\end{equation}
Here the unknown vectors $X_b$ and $Y_b$ can be taken
to be orthogonal to $u^b$. Using the explicit form \eqref{eq:amatrix} of
$A^a$ and the definitions \eqref{eq:defns}, we get the equations
\begin{subequations}
\label{eq:eigveceqns2}
\begin{align}
a X_c+\frac{1}{\rho h^*}\Big(b^b Y_b \,h_{ac}q^a -B Y_c \quad &\notag\\
+q^a h_{ac}X_7+\frac{B b_c}{\rho h} X_7\Big)&=0,\label{subeqaa}\\
B X_c-b_c \,q^aX_a- a Y_c&=0,\label{subeqbb}\\
\rho h c_s^2 q^a X_a +a X_7&=0,\label{subeqcc}\\[2pt]
\frac{p}{\rho}q^a X_a +a X_8&=0.\label{subeqdd}
\end{align}
\end{subequations}
Of course, these equations are not linearly independent when we insert
a $q^a$ corresponding to an eigenvalue.

\paragraph{Eigenvectors for the Degenerate Eigenvalues}
Setting $q^a u_a=0$ and solving Eqs.\ \eqref{eq:eigveceqns2} from the bottom
up, we get in turn $X_8$ arbitrary,
$q^a X_a=0$, $X_c=0$, $X_7=0$. Then we can take the two linearly independent
eigenvectors to have either $Y_c=0$ or $X_8=0$:
{\setlength\delimitershortfall{0pt}  %this fixes getting the bracket high enough
\begin{equation}
X_{\text{entropy}}=
\begin{bmatrix}
0\\
0\\
0\\
1
\end{bmatrix},
\qquad
X_{\text{constraint}}=\begin{bmatrix}
0\\
h_{ab}q^b\\
0\\
0
\end{bmatrix}.
\label{eq:degen}
\end{equation}}%
Here we have followed the standard convention of naming these eigenvectors the
entropy and constraint eigenvectors. The entropy eigenvector emerges
naturally when using entropy as one of the two thermodynamic
variables in the analysis, while the constraint eigenvector is associated
with the divergence constraint equation.

\paragraph{Alfv\'en Eigenvectors}

Inserting $B=-\sqrt{\rho h^*} a$ in Eqs.\ \eqref{eq:eigveceqns2}, we find
$X_8=X_7=q^aX_a=b^b Y_b=0$ and $Y_c=-\sqrt{\rho h^*}X_c$. Thus $X_c$ and
$Y_c$ are both orthogonal to $q^a$ and $b^a$. Since they must also
be orthogonal to $u^a$, we can take
\begin{equation}
X_{\text{Alf}}=
\begin{bmatrix}
\epsilon_{abcd}u^b b^c q^d\\
-\sqrt{\rho h^*}\epsilon_{abcd}u^b b^c q^d\\
0\\
0
\end{bmatrix}.
\label{eq:alf}
\end{equation}
The other eigenvector has a positive square root and a corresponding change
in $q^d$.

We can write the Alfv\'en eigenvectors in an alternative form
by introducing two
``tangential'' vectors  $\t^a_{(1,2)}$ that are orthogonal to $q^a$ and
$u^a$. (These vectors are tangential because $q^a=y_u u^a+\zeta^a$ implies
that they are spatial in the comoving frame and orthogonal to the normal
vector $\zeta^a$.) Requiring $X_c$ to be a linear combination of
these tangential vectors that is orthogonal to $b^a$ gives
for the case $B=-\sqrt{\rho h^*}a$
\begin{equation}
X_{\text{Alf}}=
\begin{bmatrix}
b_2 \t^a_{(1)}-b_1 \t^a_{(2)}\\[2pt]
-\sqrt{\rho h^*}(b_2 \t^a_{(1)}-b_1 \t^a_{(2)})\\[2pt]
0\\
0
\end{bmatrix},
\label{eq:alfalt}
\end{equation}
where $b_1= b_a \t^a_{(1)}$, $b_2=b_a\t^a_{(2)}$.
This expression agrees with that of Ref.~\cite{schoepe2018}.

\paragraph{Magnetosonic Eigenvectors}

To solve Eqs.\ \eqref{eq:eigveceqns2} in the magnetosonic case,
we can proceed as follows:
\begin{itemize}
\item
Use Eqs.\ \eqref{subeqbb} and \eqref{subeqcc} to eliminate $Y_a$ and $X_7$
from Eq.\ \eqref{subeqaa}.
\item
Dot the resulting equation with $b^c$ to get an expression for $b^c X_c$
and substitute this expression back into the equation.
\item
Solve the resulting equation for $X_c$. Simplify the result using
the eigenvalue equation \eqref{eq:N}: Solve that equation for $b^2$ and
eliminate $b^2$ from the equation for $X_c$. Discard the overall normalization
factor (which is proportional to $X^a q_a$). The result is surprisingly simple:
\begin{equation}
X_a=BGb_a-\rho h a^2 h_{ab}q^b.
\label{eq:Xsoln}
\end{equation}
\item
Use the solution \eqref{eq:Xsoln} to determine the other unknowns. The resulting
eigenvector is
\begin{equation}
X_{\text{mag}}=
\begin{bmatrix}
BGb_a-\rho h a^2 h_{ab}q^b\\[2pt]
\rho h a[(a^2+G)b_a-Bh_{ab}q^b]\\[2pt]
\rho h a\mathcal{G}\\[2pt]
pa\mathcal{G}/(\rho c_s^2)
\end{bmatrix},
\label{eq:Xmag}
\end{equation}
\end{itemize}
where, using Eq.\ \eqref{eq:N},
\begin{equation}
\mathcal{G}=[\rho h a^2(a^2+G)-B^2G]c_s^2/a^2=\rho h a^2-G b^2.
\end{equation}

\subsubsection{Left Comoving Eigenvectors}
\label{sec:leftcomoving}

The left eigenvectors follow from solving
$L(A^a q_a)=0$ analogously to the steps to find the
right eigenvectors in \S\ref{sec:comoving}. Take $L$ to be of the form
\begin{equation}
L=\begin{bmatrix}
L_b & Y_b & L_7 & L_8
\end{bmatrix},
\label{eq:Lvec}
\end{equation}
with $L_b$ and $Y_b$ 
orthogonal to $u^b$. Using the explicit form \eqref{eq:amatrix} of
$A^a$ and the definitions \eqref{eq:defns}, we get the equations
\begin{subequations}
\label{eq:leigveceqns2}
\begin{align}
a L_c-B Y_c+h_{ca}q^a(b^bY_b +\rho h c_s^2L_7+p L_8/\rho)&=0, \label{subeqlaa}\\
B L_c-b_c \,q^aL_a- \rho a h^* Y_c&=0,\label{subleqbb}\\
B\,b^aL_a+\rho h(q^aL_a+\rho a h^* L_7)&=0,\label{subleqcc}\\
a L_8&=0.\label{subleqdd}
\end{align}
\end{subequations}

\paragraph{Degenerate Eigenvectors}
There are two linearly dependent eigenvectors corresponding to the
degenerate eigenvalue given by $a=0$. In this case, we see from
Eqs.\ \eqref{subleqbb} and \eqref{subleqcc} that $L_c=0$. Then one
solution has $Y_c=0$ while the other has $L_7=L_8=0$:
\begin{equation}
\begin{split}
L_\text{entropy}&=\begin{bmatrix}
0 & 0 & -p/\rho & \rho h c_s^2
\end{bmatrix},\\
L_\text{constraint}&=\begin{bmatrix}
0 & h_{ab}q^b & 0 & 0
\end{bmatrix}.
\end{split}
\label{eq:ldegen}
\end{equation}

\paragraph{Alfv\'en Eigenvectors}
The left Alfv\'en eigenvectors are very similar to the right eigenvectors.
The $L_c$ and $Y_c$ parts are once again orthogonal to $q^a$ and $b^a$.
The eigenvector for $B=-\sqrt{\rho h^*}a$ is
\begin{equation}
L_\text{Alf}=\begin{bmatrix}
\sqrt{\rho h^*}\epsilon_{abcd}u^b b^c q^d &
-\epsilon_{abcd}u^b b^c q^d & 0 & 0
\end{bmatrix}.
\label{eq:lalf}
\end{equation}
The other eigenvector has no minus sign and the appropriate
change in $q^a$.
An alternative representation analogous to Eq.\ \eqref{eq:alfalt} is
\begin{equation}
L_\text{Alf}=
\begin{bmatrix}
\sqrt{\rho h^*}(b_2 \t^a_{(1)}-b_1 \t^a_{(2)}) &
-(b_2 \t^a_{(1)}-b_1 \t^a_{(2)}) &
0 &
0
\end{bmatrix}.
\label{eq:alfleftalt}
\end{equation}

\paragraph{Magnetosonic Eigenvectors}
By steps similar to those resulting in Eq.\ \eqref{eq:Xmag}, we find
from Eqs.\ \eqref{eq:leigveceqns2} that
the magnetosonic eigenvectors are
{\setlength\delimitershortfall{0pt}  %this gets the bracket high enough for ^T
\begin{equation}
L_\text{mag}=\left[\begin{matrix}
B(a^2+G)b_a/a-\rho h^* a h_{ab}q^b\\[2pt]
(a^2+G)b_a-Bh_{ab}q^b\\[2pt]
\mathcal{G}/(\rho h c_s^2)\\[2pt]
0
\end{matrix}
\right]^T.
\label{eq:lmagcomov}
\end{equation}}%

It is easy to check that the eigenvectors are all mutually orthogonal: The
matrix of left row eigenvectors multiplied by the matrix of the corresponding
right column vectors is diagonal (see \S IV F of Paper I).
Showing orthogonality among the four different magnetosonic eigenvectors
is actually a little more complicated and requires the explicit expressions
for the eigenvalues. The diagonal elements of the matrix product correspond
to the norms of each eigenvector. We will not bother to renormalize the
eigenvectors to unit norm, since in practice it is easier to renormalize
the eigenvectors numerically if necessary.

\subsection{Eulerian Eigensystem}
\label{sec:quasiinvert}

Most relativistic MHD codes use as primitive variables a fluid velocity and
magnetic field that are purely spatial, rather than
the pair $(u^a,b^a)$. We will use the Eulerian velocity and magnetic
field, that is, the quantities measured by an observer whose
4-velocity is $n^a$, the unit normal to the $t=\text{constant}$
hypersurface. This is the observer at rest in the $3+1$
coordinate system defined with respect to $n^a$. In
covariant form, these quantities are defined as
\begin{equation}
v^a=\gamma^{ab}u_b/W,\quad B^a= n_b \dual F^{ba}=Wb^a+u^a (n_b b^b),
\label{eq:vB}
\end{equation}
with the inverse relations
\begin{equation}
u^a=W(v^a+n^a),\qquad b^a=h^{ab}B_b/W.
\label{eq:ub}
\end{equation}
Here
\begin{equation}
W=-n_a u^a =\frac{1}{\sqrt{1-v^a v_a}}
\end{equation}
is a generalized gamma-factor and
\begin{equation}
\gamma^{ab}=g^{ab}+n^a n^b
\end{equation}
is the projection tensor orthogonal to $n^a$ that serves
as the corresponding spatial 3-metric.

\subsubsection{Transformation between $(v^a, B^a)$ and $(u^a, b^a)$}

As shown in Paper I, to get both the left and right transformed eigenvectors,
we need the Jacobian matrix for the transformation from
the 8 variables $U_\text{old}=(u^a, b^a)$
to the 6 variables
$U=(v^a, B^a)$, as well as the matrix for the inverse transformation.
Since these
Jacobian  matrices are effectively $8 \times 6$ and $6 \times 8$, they cannot
be true matrix inverses.
Recall the detailed discussion in Paper I
for the case without
a magnetic field.
We invoked a theorem
from linear algebra to conclude that, starting with the
matrix $\partial U_\text{old}/\partial U=\partial u^c/\partial v^b$,
we could find a left
inverse $\partial U/\partial U_\text{old}=\partial v^a /\partial u^c$ such that
\begin{equation}
\frac{\partial v^a}{\partial u^c}
\frac{\partial u^c}{\partial v^b}=\gamma^a{}_b.
\label{eq:inverse2}
\end{equation}
The left inverse is not unique, and we were able to
fix this non-uniqueness by imposing the additional
``inverse'' requirement
\begin{equation}
\frac{\partial u^a}{\partial v^c}
\frac{\partial v^c}{\partial u^b}=h^a{}_b.
\label{eq:inverse3}
\end{equation}
This relation plays a crucial role in showing that the
transformation is quasi-invertible, as defined
and discussed in \S\ref{I-sec:quasi}
of Paper I and also below.
The explicit Jacobian matrices satisfying these conditions are
\begin{align}
\label{eq:dudv}
\frac{\partial u^a}{\partial v^b}&=W \gamma^a{}_b + W^2 u^a v_b,\\
\frac{\partial v^a}{\partial u^b}
&=\frac{1}{W}\delta^a{}_b+\frac{1}{W^2}u^a n_b.
\label{eq:vtrans}
\end{align}
Formally, we have satisfied the quasi-invertible condition of
Paper I Eq.\ \eqref{I-eq:quasiinvertible},
\begin{equation}
A^a_{\rm old}=A^a\frac{\partial U}{\partial U_{\rm old}}
\quad \Leftrightarrow\quad
A^a=A^a_{\rm old}\frac{\partial U_{\rm old}}{\partial U},
\label{eq:quasii}
\end{equation}
by the conditions
\begin{align}
\label{eq:leftinverse}
\frac{\partial U}{\partial U_{\rm old}}
\frac{\partial U_{\rm old}}{\partial U}&=\one,\\
\frac{\partial U_{\rm old}}{\partial U}
\frac{\partial U}{\partial U_{\rm old}}&=\one.
\label{eq:makeunique}
\end{align}

We now need to generalize these transformations to include
the magnetic field.
We will see that, while we are always guaranteed to be able
to find a left inverse $\partial U/\partial U_{\rm old}$
for $\partial U_{\rm old}/\partial U$ as in Eq.\ \eqref{eq:leftinverse},
we will not be able to make this left inverse unique by imposing
Eq.\ \eqref{eq:makeunique}. Fortunately, we can find an alternative
criterion that still gives us a quasi-invertible transformation.
To see how to proceed, we will attempt to enforce Eq.\ \eqref{eq:makeunique}
and see where it breaks down.

We start by differentiating $U_{\rm old}=(u^a,b^a)$.
Since in Eqs.\ \eqref{eq:vB} and
\eqref{eq:ub} the relations between $u^a$ and $v^a$ are independent
of the magnetic field, Eqs.\ \eqref{eq:dudv} and \eqref{eq:vtrans}
are still valid. In addition, contributions to the Jacobian
matrices like $\partial v^a /\partial b^b$ and $\partial u^a/\partial B^b$
are zero. To find the non-zero terms,
start with Eq.\ \eqref{eq:ub} in the form
\begin{equation}
 b^a=h^a{}_c \gamma^c{}_bB^b/W.
\end{equation}
Differentiating gives
\begin{equation}
\frac{\partial b^a}{\partial B^b}=\frac{1}{W}h^a{}_c \gamma^c{}_b.
\label{eq:dbdB}
\end{equation}
Similarly, since $\partial W/\partial v^b=W^3 v_b$, we get
\begin{align}
\frac{\partial b^a}{\partial v^b}
&= -Wv_b h^a{}_c B^c+\frac{1}{W}\left(\frac{\partial u^a}{\partial v^b}u_c
+\frac{\partial u_c}{\partial v^b}u^a\right)B^c\notag\\
&=u^a B_b-WB^a v_b +W B^c v_c(\gamma^a{}_b+Wu^a v_b),
\label{eq:dbdv}
\end{align}
where we have used Eq.\ \eqref{eq:dudv} and made some simple substitutions.

Now Eq.\ \eqref{eq:leftinverse} becomes (cf.\ \eqref{eq:inverse2} with
no magnetic field)
\begin{align}
\frac{\partial(v^a,B^a)}{\partial (u^c,b^c)}
\frac{\partial(u^c,b^c)}{\partial (v^b,B^b)}&=
\begin{bmatrix}
\dfrac{\partial v^a}{\partial u^c} & 0\\[7pt]
\dfrac{\partial B^a}{\partial u^c} &
\dfrac{\partial B^a}{\partial b^c}
\end{bmatrix}
\begin{bmatrix}
\dfrac{\partial u^c}{\partial v^b} & 0\\[7pt]
\dfrac{\partial b^c}{\partial v^b} & 
\dfrac{\partial b^c}{\partial B^b}
\end{bmatrix}\notag\\[2pt]
&=\begin{bmatrix}
\gamma^a{}_b & 0\\
0 & \gamma^a{}_b
\end{bmatrix}.
\label{eq:jacobian}
\end{align}
This gives two conditions on the magnetic field part of the
transformation:
\begin{align}
\label{eq:first}
\dfrac{\partial B^a}{\partial u^c}\dfrac{\partial u^c}{\partial v^b}
+\dfrac{\partial B^a}{\partial b^c}\dfrac{\partial b^c}{\partial v^b}
&=0,\\
\dfrac{\partial B^a}{\partial b^c}\dfrac{\partial b^c}{\partial B^b}
&=\gamma^a{}_b.
\label{eq:second}
\end{align}

Let's first consider Eq.\ \eqref{eq:second} and determine $\partial B^a/
\partial b_c$. The most general expression for this quantity can
be written in term of $n^a$, $u^a$, $B^a$, and the metric. Also,
from Eq.\ \eqref{eq:vB} we expect the magnetic field to enter the Jacobian
matrix at most linearly. So we set
\begin{multline}
\dfrac{\partial B^a}{\partial b^c}=
A \delta^a{}_c +B u^a u_c +C u^a n_c +D n^a u_c
+E n^a n_c\\
+F B^a u_c +G u^a B_c +H B^a n_c +I n^a B_c.
\end{multline}
(Note that the scalars $A$, $B$, \dots used in this subsection are not
related to the same symbols used elsewhere in the paper.)
Substitute this expression and Eq.\ \eqref{eq:dbdB} in Eq.\ \eqref{eq:second}.
Thus find
\begin{equation}
\dfrac{\partial B^a}{\partial b^c}=
W \delta^a{}_c +B u^a u_c + u^a n_c +D n^a u_c+F B^a u_c.
\label{eq:dBdbpartial}
\end{equation}

We'll return to Eq,\ \eqref{eq:first} in a moment.
Note that the ``inverse'' requirement \eqref{eq:makeunique}
multiplies the matrices in Eq.\ \eqref{eq:jacobian} in the opposite order.
This gives the two conditions
\begin{align}
\label{eq:invfirst}
\dfrac{\partial b^a}{\partial v^c}
\dfrac{\partial v^c}{\partial u^b}
+
\dfrac{\partial b^a}{\partial B^c}
\dfrac{\partial B^c}{\partial u^b}
&\stackrel{?}{=}0,\\
\dfrac{\partial b^a}{\partial B^c}
\dfrac{\partial B^c}{\partial b^b}
&=h^a{}_b.
\label{eq:invsecond}
\end{align}
The question mark over the equals sign in Eq.\ \eqref{eq:invfirst} is
to remind us that we will find that we
cannot in fact accomplish this requirement.

Turn attention first to
Eq.\ \eqref{eq:invsecond}. Using Eqs.\ \eqref{eq:dbdB} and 
\eqref{eq:dBdbpartial}, we find $B=F=0$ in Eq.\ \eqref{eq:dBdbpartial}.

Next, set
\begin{multline}
\dfrac{\partial B^a}{\partial u^c}=
\bar A \delta^a{}_c +\bar B u^a u_c +\bar C u^a n_c +\bar D n^a u_c
+\bar E n^a n_c\\
+\bar F B^a u_c +\bar G u^a B_c +\bar H B^a n_c +\bar I n^a B_c.
\end{multline}
Substitute this expression into Eq.\ \eqref{eq:invfirst} and
choose the coefficients to zero the expression. However, \emph{do
not try to zero the terms proportional to} $u^a$. You will find
that it is not possible
to do so and simultaneously zero all the other terms. Instead, use
the corresponding coefficients to help zero the remaining terms.
We will explain at the end of this subsection why this procedure works.

Proceeding in this way, we find
\begin{multline}
\dfrac{\partial B^a}{\partial u^c}=\bar D n^a u_c+\bar E n^a n_c
+\bar I n^a B_c\\
+\frac{1}{W}[B^a(Wu_c-n_c)
-(B^b u_b)(W\delta^a{}_c+u^a n_c)].
\end{multline}
Finally, substitute the results so far into Eq.\ \eqref{eq:first}. Thus
find $D = W \bar I$ and $\bar E=0$.

We are free to set the remaining coefficients $\bar D$, $\bar I$ and $D$ to zero
so that
\begin{equation}
\frac{\partial}{\partial u^b}(B^a n_a)=
\frac{\partial}{\partial b^b}(B^a n_a)
=0.
\end{equation}
Thus the final form of the transformation matrices is
\begin{align}
\label{eq:dBdufinal}
\dfrac{\partial B^a}{\partial u^b}&=
\frac{1}{W}[B^a(Wu_b-n_b)
-(B^b u_b)(W\delta^a{}_b+u^a n_b)],\\
\dfrac{\partial B^a}{\partial b^b}&=
W \delta^a{}_b + u^a n_b.
\label{eq:dBdbfinal}
\end{align}

We now explain why the above procedure works.
Recall that a quasi-invertible transformation
ensures that the matrices of right and left eigenvectors are inverses:
\begin{equation}
\one = L\cdot X =L_{\rm old}\left(A^a_{\rm old}n_a\right)
\frac{\partial U_{\rm old}}{\partial U}
\frac{\partial U}{\partial U_{\rm old}}X_{\rm old}.
\label{eq:transgivesinverses}
\end{equation}
Here we have used the transformation rules for going from the comoving
to the Eulerian eigenvectors, 
Eqs.\ \eqref{I-eq:transeigen} and \eqref{I-eq:ltrans} of Paper I.
Notice that the factor
$\left(A^a_{\rm old}n_a\right)$ gives zero when acting on $u^a$.
Accordingly, we can allow the right-hand side of Eq.\ \eqref{eq:makeunique}
to differ from the identity by an off-diagonal term proportional to $u^a$ and
Eq.\ \eqref{eq:transgivesinverses} will still hold.
(In fact, 
instead of zero, the right-hand side of Eq.\ \eqref{eq:invfirst} turns out to
be $u^a b_b$.) Since the inverse of a matrix is unique, we are assured
that any procedure that finds the correct inverse $L$ of the matrix $X$
must be valid.
Essentially what we have done is to require that the
quasi-invertible condition \eqref{eq:quasii} only needs to hold
when projected along $n_a$.

\subsubsection{Eigenvalues for the Eulerian frame}

In the Eulerian frame, the vector $q_a$ defined
in Eq.~\eqref{eq:qu} becomes
\begin{equation}
q_a=y n_a+s_a,
\end{equation}
where $s_a$ is a unit spatial vector in the direction of the decomposition.
Here $y$ is now the eigenvalue in the Eulerian frame.

The degenerate eigenvalues corresponding to $a=q^a u_a=0$ have the value in
the Eulerian frame
\begin{equation}
y=v^a s_a.
\end{equation}
For the Alfv\'en eigenvalues, the condition
$B=-\sqrt{\rho h^*}a$ gives
\begin{equation}
y_\text{Alf}=v^a s_a+\frac{B^as_a}{W^2(\sqrt{\rho h^*}+B^a v_a)}.
\label{eq:alfeuler}
\end{equation}
The other eigenvector has the opposite sign for the square root.
The magnetosonic eigenvalues require the solution of the quartic
equation \eqref{eq:N} expressed in Eulerian variables.

\subsubsection{Right Eigenvectors in the Eulerian frame}

The transformation from the comoving to the Eulerian right eigenvectors
is given by the Jacobian matrix $\partial(v^a,B^a)/\partial(u^b,b^b)$.
This matrix acts on the vector part of each eigenvector. The scalar
part of the eigenvectors is unchanged.

\paragraph{Degenerate Eigenvectors}
Applying this transformation to the degenerate eigenvectors
\eqref{eq:degen} and setting $q^a=y n^a+s^a$, we get
{\setlength\delimitershortfall{0pt}  %this fixes getting the bracket high enough
\begin{equation}
\xbf_{\text{entropy}}=
\begin{bmatrix}
0\\
0\\
0\\
1
\end{bmatrix},
\quad
\xbf_{\text{constraint}}=\begin{bmatrix}
0\\
s^a-(s^b v_b)v^a\\
0\\
0
\end{bmatrix}.
\label{eq:xfontent}
\end{equation}}%

\paragraph{Alfv\'en Eigenvectors}
For the Alfv\'en eigenvectors, we have not found a simple way of
transforming the expression \eqref{eq:alfalt}.
Thus we start
with Eq.\ \eqref{eq:alf}.
The transformation of the first component of that
expression consists of the following
steps.
\begin{itemize}
\item
The Jacobian matrix \eqref{eq:vtrans} gives
\begin{equation}
\frac{\partial v^a}{\partial u^e}\epsilon^e{}_{bcd}u^b b^c q^d
=\frac{1}{W}\epsilon^a{}_{bcd}u^b b^c q^d +\frac{1}{W^2}
u^a n_e\epsilon^e{}_{bcd}u^b b^c q^d.
\end{equation}
\item
Replace $b^a$ and $u^a$ by their Eulerian equivalents $B^a$ and $v^a$,
Eq.\ \eqref{eq:ub}. Also set  $q^a=y_\text{Alf} n^a+s^a$.
\item
Expand $B^a$ and $v^a$ in the orthonormal Eulerian spatial basis:
\begin{equation}
\begin{split}
B^a&=B_n s^a+B_1 t^a_{(1)}+B_2 t^a_{(2)},\\
v^a&=v_n s^a+v_1 t^a_{(1)}+v_2 t^a_{(2)}.
\end{split}
\end{equation}
The $(s^a,t^a_{(1)}, t^a_{(2)})$ components of the result are all
proportional to $\epsilon_{abcd}n^a s^b t^c_{(1)}, t^d_{(2)}$, which
is unity. 
\item
Substitute for $y_\text{Alf}$ from Eq.\ \eqref{eq:alfeuler}.
Simplify the result using $B_n v_n+B_1 v_1+B_2 v_2=B^a v_a$
and $v_n^2+v_1^2+v_2^2=1-1/W^2$.
\end{itemize}
Repeat the procedure to find the second component of the eigenvector
using Eqs.\ \eqref{eq:dBdufinal} and \eqref{eq:dBdbfinal}. The result is
\begin{equation}
\xbf_\text{Alf}=\begin{bmatrix}
B_n(B_1v_2-B_2 v_1)/r_1 W^2\\
(r_2B_2-r_3v_2)/r_1 W^2\\
(-r_2B_1+r_3v_1)/r_1 W^2\\
0\\
r_4 v_2+B_2\sqrt{\rho h^*}\\
-r_4 v_1-B_1\sqrt{\rho h^*}\\
0\\
0
\end{bmatrix},
\label{eq:alffont}
\end{equation}
where
\begin{equation}
\begin{alignedat}{2}
\label{eq:r1}
r_1&=B^a v_a+\sqrt{\rho h^*},&\quad r_2&=B_n v_n-r_1,\\
r_3&=B_n^2+r_1B^a v_a W^2,&\quad r_4&=B^aB_a+r_1B^a v_a W^2.
\end{alignedat}
\end{equation}
In Eq.\ \eqref{eq:alffont} the first 6 components are the
$(s^a,t^a_{(1)}, t^a_{(2)})$ components of each of the vectors.

Equation (22) in Ref.~\cite{ibanez2015} gives an implicit expression
for $\xbf_\text{Alf}$
that uses $(B^a,v^a,s^a)$ as a basis instead of the orthonormal
basis $(s^a,t^a_{(1)}, t^a_{(2)})$. We have computed that expression
explicitly and have verified that our expression
is equivalent up to the normalization, but is somewhat
simpler.\footnote{There is a typo in the (2,2) term of the
characteristic matrix in~\cite{ibanez2015}.
The term $B^iB_j$ should be $-B^iB_j$.}

\paragraph{Magnetosonic Eigenvectors}
Next we consider the magnetosonic eigenvectors. We need to apply the Jacobian
matrix $\partial(v^a,B^a)/\partial(u^b,b^b)$ to the expression
\eqref{eq:Xmag}. For the first component of Eq.\ \eqref{eq:Xmag}, we find
that
\begin{multline}
\frac{\partial v^a}{\partial u^c}(BGb^c-\rho h a^2 h^{cb}q_b)\\
=
\frac{BG(b^bn_b u^a+Wb^a)-\rho h a^2(n^b q_bu^a+W q^a)}{W^2}.
\end{multline}
Replace $b^a$ and $u^a$ by their Eulerian equivalents $B^a$ and $v^a$,
Eq.\ \eqref{eq:ub}. Also set  $q^a=y_\text{mag} n^a+s^a$.
Thus get
\begin{equation}
\frac{BG B^a -\rho h a^2(s^a-y_\text{mag}v^a)}{W^2}.
\end{equation}
Transforming the second component of Eq.\ \eqref{eq:Xmag} in the same way
with Eqs.\ \eqref{eq:dBdufinal} and \eqref{eq:dBdbfinal},
we get
\begin{equation}
\frac{\rho h a}{W}[B^a(GW-y_\text{mag}a)-W^2(B-a B^b v_b)(s^a-y_\text{mag} v^a)].
\end{equation}
Making the replacement $B=a B^b v_b+B^b s_b/W$ gives the final
expression
\begin{equation}
\xbf_\text{mag}=
\begin{bmatrix}
\dfrac{BG B^a -\rho h a^2W(s^a-y_\text{mag}v^a)}{W^2}\\[3pt]
\rho h a [B^a(1-y_\text{mag} v_n)-B_n(s^a-y_\text{mag} v^a)]\\[2pt]
\rho h a\mathcal{G}\\[2pt]
pa\mathcal{G}/(\rho c_s^2)
\end{bmatrix}.
\label{eq:xmag}
\end{equation}
The above equation  agrees with Eq.~(25) in \cite{ibanez2015}.
They remark, ``The expressions of the eigenvectors have been obtained after
tedious algebraic manipulations.'' And this for only the right
eigenvectors, and not in terms of the conserved variables.

We can also express the vector parts of Eq.\ \eqref{eq:xmag} in
the $(s^a,t^a_{(1)}, t^a_{(2)})$ basis to give the result
\begin{equation}
\xbf_\text{mag}=
\begin{bmatrix}
[BGB_n-\rho h a^2W (1-y_\text{mag}v_n)]/W^2\\[2pt]
[BGB_1+\rho h a^2W y_\text{mag}v_1]/W^2\\[2pt]
[BGB_2+\rho h a^2W y_\text{mag}v_2]/W^2\\
0\\
\rho h a[B_1(1-y_\text{mag}v_n)+v_1B_ny_\text{mag}]\\[2pt]
\rho h a[B_2(1-y_\text{mag}v_n)+v_2B_ny_\text{mag}]\\[2pt]
\rho h a\mathcal{G}\\[2pt]
pa\mathcal{G}/(\rho c_s^2)
\end{bmatrix}.
\label{eq:xmagfinal}
\end{equation}

\subsubsection{Left Eigenvectors in the Eulerian frame}
\label{sec:lefteulerian}

The transformation from comoving to Eulerian frames
for the left eigenvectors is more complicated than
the one for the right eigenvectors. The reason is that the transformation
is not invertible, but only quasi-invertible. As shown in Paper I,
the transformation is given by
\begin{equation}
L=L_{\rm old}\left(A^a_{\rm old}n_a\right)\frac{\partial U_{\rm old}}{\partial U}
\label{eq:ltrans}
\end{equation}
(Eq.\ \ref{I-eq:ltrans} of Paper I). Here the matrix
matrix $\partial U_{\rm old}/\partial U$  must be chosen to satisfy the
quasi-invertible condition and takes the form
\begin{equation}
\frac{\partial U_{\rm old}}{\partial U}=
\begin{bmatrix}
\dfrac{\partial u^c}{\partial v^b} & 0 & 0 & 0\\[7pt]
\dfrac{\partial b^c}{\partial v^b} &
\dfrac{\partial b^c}{\partial B^b} & 0 & 0\\[7pt]
0 & 0 & 1 & 0\\
0 & 0 & 0 & 1
\end{bmatrix}.
\end{equation}
Using Eqs.\ \eqref{eq:amatrix}, \eqref{eq:dudv}, \eqref{eq:dbdB}, and
\eqref{eq:dbdv}, we find
after some simplifications
\begin{widetext}
\begin{equation}
\left(A^a_{\rm old}n_a\right)\frac{\partial U_{\rm old}}{\partial U}
=
\begin{bmatrix}
C^{ab}
&
-\dfrac{1}{\rho h^* W}[(b^cn_c)h^{ab}+WH^ab^b]
&\dfrac{1}{\rho^2 h h^*}[(b^cn_c)b^a-\rho h W H^a] & 0\\
0 & -(\gamma^{ab}+W u^av^b) & 0 & 0\\
-\rho h c_s^2 W^3 v^b & 0 & -W & 0\\
-\dfrac{p}{\rho}W^3 v^b & 0 & 0 & -W
\end{bmatrix}.
\label{eq:a0}
\end{equation}
\end{widetext}
Here
\begin{equation}
\begin{split}
C^{ab}&=
\dfrac{1}{\rho h^*}\{[(b^cn_c)^2-\rho h^* W^2]h^{ab}+W (b^cn_c) H^a b^b\\
&\quad+W^3H^a(b^2v^b+\rho h^* n^b)
+W^2(b^cn_c)b^av^b\},\\
H^a&=u^a-n^a/W.
\end{split}
\end{equation}

\paragraph{Degenerate Eigenvectors}
Applying the transformation \eqref{eq:a0} to the eigenvectors \eqref{eq:ldegen}
corresponding to the degenerate eigenvalues gives
\begin{equation}
\begin{split}
\lbf_\text{entropy}&=\begin{bmatrix}
0 & 0 & -p/\rho & \rho h c_s^2
\end{bmatrix},\\
\lbf_\text{constraint}&=\begin{bmatrix}
0 & \gamma_{ab}q^b & 0 & 0
\end{bmatrix}.
\end{split}
\label{eq:ldegeneuler}
\end{equation}

\paragraph{Alfv\'en Eigenvectors}
Applying the transformation \eqref{eq:a0} to the Alfv\'en eigenvectors 
\eqref{eq:lalf} is straightforward but a little lengthy.
We have verified our calculations with computer algebra.
The transformation consists of the following steps.
\begin{itemize}
\item
Multiply the eigenvector \eqref{eq:lalf} into the matrix \eqref{eq:a0}
and then convert all quantities to their Eulerian counterparts using
Eq.\ \eqref{eq:ub}.
\item
Substitute $q^a=y_\text{Alf}n^a+s^a$ with $y_\text{Alf}$ from
Eq.~\eqref{eq:alfeuler}.
\item
Expand the vectors $B^a$ and $v^a$ when contracted into $\epsilon_{abcd}$
in the $(s^a,t^a_{(1)}, t^a_{(2)})$ basis.
\item
Project the vector pieces of the eigenvector along $(s^a,t^a_{(1)},
t^a_{(2)})$ and simplify. A useful relation for this is
\begin{equation}
B^a B_a=[\rho h^* -(B^av_a)^2]W^2-\rho h W^2.
\end{equation}
\item
Divide out a common scale factor of
$\epsilon_{abcd}n^a s^b t^c_{(1)}t^d_{(2)}$.
\end{itemize}
The result is
\begin{equation}
\lbf_\text{Alf} =
\left[\begin{matrix}
l_1(B_2v_1-B_1v_2)\\
l_1(B_n v_2-B_2 v_n)+B_2(b^2+\rho h W^2)\\
l_1(B_1 v_n-B_n v_1)-B_1(b^2+\rho h W^2)\\
\sqrt{\rho h^*} v_n(B_1 v_2-B_2 v_1)\\
-\sqrt{\rho h^*} B_2(1-v_n^2)-v_2(\sqrt{\rho h^*}B_n v_n+b^2)\\
\sqrt{\rho h^*} B_1(1-v_n^2)+v_1(\sqrt{\rho h^*}B_n v_n+b^2)\\
B_2 v_1-B_1 v_2\\
0
\end{matrix}\right]^T,
\label{eq:lalffinal}
\end{equation}
where
\begin{equation}
l_1=B_n\sqrt{\rho h^*}-\rho h v_n W^2.
\end{equation}

\paragraph{Magnetosonic Eigenvectors}
For the magnetosonic eigenvectors, we multiply Eq.\ \eqref{eq:lmagcomov}
into Eq.\ \eqref{eq:a0}. We use the definitions \eqref{eq:defns} and
then convert the remaining quantities to their Eulerian counterparts
using Eq.\ \eqref{eq:ub}. Simplify the eigenvector using the eigenvalue
equation \eqref{eq:N} to replace a term $B^2G$.
The manipulations
are straightforward but a little lengthy,
and we have again verified the results with computer algebra.
The result is
\begin{equation}
\lbf_\text{mag} =
\left[\begin{matrix}
a(B^aB_a+\rho h W^2)s_a-(S v_a+TB_a)/a\\[3pt]
(aB_n+KB^bv_b)v_a+KB_a/W^2+B_ns_a/W\\[4pt]
K+\dfrac{GBB_n}{\rho h a^2}\\[8pt]
0
\end{matrix}
\right]^T,
\label{eq:lmagfinal}
\end{equation}
where
\begin{equation}
\begin{split}
K&=ay_\text{mag}-GW=-W(1-v_n y_\text{mag}),\\
S&=B_n GW^2 B^av_a+aK B^aB_a,\\
T&=B_n(a^2+G)-aKB^av_a.
\end{split}
\end{equation}

\subsubsection{Transformation to $(\rho,\epsilon)$}
\label{sec:rhoeps}

Since many numerical codes use the pair
$(\rho,\epsilon)$ as primitive variables rather than $(p,\epsilon)$,
we transform the Eulerian eigenvectors determined above to
these variables.
(The transformation procedure is similar for any other pair of
thermodynamic variables.)
The transformation is invertible:
\begin{align}
\label{eq:kappa}
\frac{dU_{\rm old}}{dU}&=
\frac{\partial(p,\epsilon)}{\partial(\rho,\epsilon)}
=\begin{bmatrix}
\chi & \kappa\\
0 & 1
\end{bmatrix},\\
\frac{dU}{dU_{\rm old}}&=
\frac{\partial(\rho,\epsilon)}{\partial(p,\epsilon)}
=\begin{bmatrix}
1/\chi & -\kappa/\chi\\
0 & 1
\end{bmatrix}.
\label{eq:inversekappa}
\end{align}
The part of the transformation acting on the vector part of the eigenvectors
is the identity matrix, so we have not bothered to write it explicitly.
Applying Eq.\ \eqref{eq:inversekappa} to the Eulerian right eigenvectors above,
we find that $\xbf_{\text{constraint}}$ and $\xbf_{\text{Alf}}$ are unchanged,
while
\begin{equation}
\begin{gathered}
\xbf_{\text{entropy}}=
\begin{bmatrix}
0\\
0\\
-\kappa\\
\chi
\end{bmatrix},\\[4pt]
\xbf_\text{mag}=
\begin{bmatrix}
\dfrac{BG B^a -\rho h a^2W(s^a-y_\text{mag}v^a)}{W^2}\\[3pt]
\rho h a [B^a(1-y_\text{mag} s^b v_b)-B^b s_b(s^a-y_\text{mag} v^a)]\\[2pt]
\rho a\mathcal{G}/c_s^2\\[2pt]
pa\mathcal{G}/(\rho c_s^2)
\end{bmatrix}.
\end{gathered}
\label{eq:xmag2}
\end{equation}
Note that we have not bothered to introduce a separate
notation to distinguish $(p,\epsilon)$ eigenvectors from $(\rho,\epsilon)$
eigenvectors since these are only being used as intermediate
quantities on the way to using conserved variables.

Similarly, applying Eq.\ \eqref{eq:kappa} to the left eigenvectors leaves
$\lbf_\text{constraint}$ unchanged. The entropy eigenvector becomes
\begin{equation}
\lbf_\text{entropy}=\begin{bmatrix}
0 & 0 & -\chi p/\rho & \chi \rho
\end{bmatrix}.
\end{equation}
Equation \eqref{eq:lalffinal} for $\lbf_\text{Alf}$ is unchanged, except that
the last two elements become
\begin{equation}
\left[
\begin{matrix}
\chi (B_2 v_1-B_1 v_2)\\
\kappa (B_2 v_1-B_1 v_2)
\end{matrix}
\right]^T.
\label{eq:lalf_rhoeps}
\end{equation}
Equation \eqref{eq:lmagfinal}
for $\lbf_\text{mag}$ is unchanged, except that the last two elements become
\begin{equation}
\left[
\begin{matrix}
\chi(K+GBB_n/\rho h a^2)\\
\kappa(K+GBB_n/\rho h a^2)
\end{matrix}
\right]^T.
\label{eq:lmag_last_two}
\end{equation}

\section{Conserved Formulation}
\label{sec:consvariables}

In this section we introduce the conserved variables, which
are almost universally used in practical codes. Our ultimate goal is to
give the complete characteristic decomposition for the evolution of these
variables. However, this decomposition \emph{cannot} be found by simply
transforming our previous results to these new variables. The reason
is that as shown in Ref.~\cite{schoepe2018}, the formulation of GRMHD
in terms of conservation equations is \emph{not} simply a linear combination
of the Anile evolution equations. The two formulations differ by
multiples of the divergence constraint. Since the divergence constraint
contains derivatives, the principal parts of the equations differ, and hence
so do the eigensystems.

Ref.~\cite{schoepe2018} also showed that
the popular ``Valencia'' conservation formulation of GRMHD is only weakly
hyperbolic: For certain directions of the spatial normal chosen for making
the characteristic decomposition, there is not a complete set of eigenvectors.
This implies that the system is not well posed and is not a good candidate
for numerical simulations. Accordingly, in the next section we will focus
on a divergence-cleaning conservation formulation that is well posed.

In the remainder of this section we will define the usual set of conserved
variables. We then give the transformation matrix between these variables
and the primitive variables, and also give the inverse matrix.
Next we transform the Anile eigenvectors to the conserved variables.
We must emphasize again that these eigenvectors are not the eigenvectors
for the evolution system of the conserved variables. We present them
to make contact with previous literature, and
also because the actual eigenvectors
we will want in the end are very similar to these ones.

In Appendix \ref{app:anton} we discuss the relation between our approach
and the method of Ref.~\cite{anton2010}.

\subsection{Conserved Variables}

The primitive variables in the Eulerian frame are
\begin{equation}
  U_\text{old}= [v^a, B^a, \rho, \epsilon].
\label{eq:primitive}
\end{equation}
We will want to transform from these variables to the conserved variables
\begin{equation}
\begin{split}
S^a &=\rho h^* W^2 v^a-(\alpha b^0)b^a,\\
B^a &= B^a,\\
D &=\rho W,\\
\tau &= \rho h^* W^2 -p^*-(\alpha b^0)^2 - \rho W.
\end{split}
\end{equation}
Here
\begin{equation}
\alpha b^0 =-n_a b^a=u_a B^a= W (v_i B^i).
\end{equation}
Alternative expressions in terms of the Eulerian magnetic field are
\begin{equation}
\begin{split}
S^a&=(\rho h W^2+B^b B_b)v^a-(B^b v_b)B^a,\\
\tau &= \rho h W^2-p-\rho W+\left(1-\frac{1}{2W^2}\right)B^a B_a
- \tfrac{1}{2}(B^av_a)^2.
\end{split}
\label{eq:consalt}
\end{equation}

\subsection{Transformation Matrix for Right Eigenvectors}
\label{sec:transmatright}

We will get the eigenvectors for the conservative formulation by transforming
from a primitive variable formulation rather than deriving them
directly from the equations of motion in conservation form.
Using $U_\text{old}$ from Eq.~\eqref{eq:primitive}, and denoting the
conserved variables by $U=(S^a,B^a,D,\tau)$,
the transformation matrix
$\partial U/\partial U_{\rm old}$
consists of the following derivatives:
\begin{widetext}
\begin{multline}
\frac{\partial(S^a,B^a,D,\tau)}{\partial(v^b,B^b,\rho,\epsilon)}=\\
\begin{bmatrix}
\rho h W^2(\gamma^a{}_b+2W^2v^a v_b)+(B^cB_c)\gamma^a{}_b -B^aB_b &
   2v^aB_b-B^av_b-(v^cB_c)\gamma^a{}_b & 
  (1+\epsilon+\chi)W^2 v^a &
  (\rho+\kappa)W^2 v^a\\[2pt]
0 & \gamma^a{}_b & 0 & 0\\
\rho W^3 v_b &
0 &
W &
0\\[2pt]
\rho (2h W-1) W^3 v_b +(B^cB_c)v_b-(v^cB_c)B_b&
(2-1/W^2)B_b-(v^cB_c)v_b &
(1+\epsilon+\chi)W^2-\chi-W &
(\rho+\kappa)W^2-\kappa
\end{bmatrix}.\\
\label{eq:transtocons}
\end{multline}
\end{widetext}
Here we have used the expressions \eqref{eq:consalt}, and
the derivatives of $\rho h$ have been
computed using $\rho h = \rho+\rho\epsilon+p$.
It is convenient when using this matrix to replace $1+\epsilon$
by $h-p/\rho$ since this makes it easier to
introduce the sound speed \eqref{eq:cs}.

\subsection{Transformation Matrix for Left Eigenvectors}
\label{sec:transmatleft}

We need the inverse of the Jacobian matrix \eqref{eq:transtocons} to compute
the left eigenvectors. 
The inverse can be found by computing
the necessary partial derivatives directly.
We use the idea in Refs.\ \cite{antonthesis,anton2010}
of defining an
intermediate variable
\begin{equation}
Z=\rho h W^2.
\label{eq:zdef}
\end{equation}
Write the primitive variables in terms of the conserved variables and
$Z$:
\begin{align}
\rho&=\frac{D}{W},\label{eq:rhoprim}\\
\epsilon&=-1+\frac{Z}{DW}-\frac{pW}{D},\label{eq:epsprim}\\
p&=Z-\tau-D-\frac{(B^a S_a)^2}{2Z^2}+\left(1-\frac{1}{2W^2}\right)B^a B_a,
  \label{eq:pprim}\\
v^a&=\frac{S^a}{Z+B^a B_a}+\frac{B^a(B^b S_b)}{Z(Z+B^a B_a)}. \label{eq:vprim}
\end{align}
The last two equations follow from Eqs.~\eqref{eq:consalt}.

As an example of the calculations, consider the derivatives of the
primitive variables $(v^a, B^a, \rho, \epsilon)$ with respect to
$\tau$, keeping $S^a$, $B^a$, and $D$ constant. Differentiating
Eq.~\eqref{eq:vprim} and simplifying with the first equation in
\eqref{eq:consalt}, we get
\begin{equation}
\frac{\partial v^a}{\partial \tau}= -\frac{B^a(B^b S_b)+Z^2 v^a}{Z^2 (Z+
B^a B_a)}\frac{\partial Z}{\partial \tau}.
\label{eq:dvdtau}
\end{equation}
Since $\partial W/\partial v^a = W^3 v_a$, we find
\begin{equation}
\frac{\partial W}{\partial \tau}=\frac{W[Z(1-W^2)-W^2(B^a v_a)^2]}
  {Z(Z+B^aB_a)}\frac{\partial Z}{\partial \tau}.
\label{eq:dwdtau}
\end{equation}
Next, Eq.~\eqref{eq:rhoprim} gives
\begin{equation}
\frac{\partial \rho}{\partial \tau}=-\frac{D}{W^2}
  \frac{\partial W}{\partial \tau}.
\label{eq:drhodtau}
\end{equation}
Eq.~\eqref{eq:pprim} gives, after some simplifications,
\begin{equation}
\frac{\partial p}{\partial \tau}=-1+\frac{Z+b^2}{Z+B^a B_a}
  \frac{\partial Z}{\partial \tau}.
\label{eq:dpdtau}
\end{equation}
Then Eq.~\eqref{eq:epsprim} gives after simplification
\begin{equation}
\frac{\partial \epsilon}{\partial \tau}= \frac{W}{D}-\frac{p}{D}
  \frac{\partial W}{\partial \tau}.
\end{equation}
The final relation we need is one for $\partial Z/\partial \tau$.
We get this by using the equation of state information, $p(\rho,\epsilon)$:
\begin{equation}
\frac{\partial p}{\partial \tau}=\chi \frac{\partial \rho}{\partial \tau}
+\kappa \frac{\partial \epsilon}{\partial \tau}.
\end{equation}
Equating this expression to Eq.~\eqref{eq:dpdtau} and solving the
resulting linear equation for $\partial Z/\partial \tau$ gives
\begin{equation}
\frac{\partial Z}{\partial \tau} = \frac{(\kappa+\rho)(Z+B^aB_a)}
{\mathcal{D}},
\label{eq:dzdtau}
\end{equation}
where
\begin{equation}
\mathcal{D}=\rho[b^2-c_s^2 (B^a v_a)^2+\rho h c_s^2 +Z(1-c_s^2)].
\end{equation}
Of course, $\partial B^a/\partial \tau=0$, as are $\partial B^a/\partial D$
and $\partial B^a/\partial S^b$.

Using a similar procedure for the other variables, we find
{\allowdisplaybreaks
\begin{align}
\frac{\partial v^a}{\partial D}&= -\frac{B^a(B^b S_b)+Z^2 v^a}{Z^2 (Z+
B^a B_a)}\frac{\partial Z}{\partial D},\\
\frac{\partial W}{\partial D}&=\frac{W[Z(1-W^2)-W^2(B^a v_a)^2]}
  {Z(Z+B^aB_a)}\frac{\partial Z}{\partial D},\\
\frac{\partial \rho}{\partial D}&=\frac{1}{W}-\frac{D}{W^2}
  \frac{\partial W}{\partial D},\\
\frac{\partial p}{\partial D}&=-1+\frac{Z+b^2}{Z+B^a B_a}
  \frac{\partial Z}{\partial D},\\
\frac{\partial \epsilon}{\partial D}&= \frac{(D+p)W^2-Z}{D^2W}-\frac{p}{D}
  \frac{\partial W}{\partial D},\\
\frac{\partial Z}{\partial D} &= \frac{[(\kappa(W-h)+\rho h c_s^2
+\rho W](Z+B^aB_a)}
{W\mathcal{D}},\\
\frac{\partial v^a}{\partial S^b}&=
\frac{Z\gamma^a{}_b+B^a B_b}{Z(Z+B^a B_a)}
 -\frac{B^a(B^b S_b)+Z^2 v^a}{Z^2 (Z+B^a B_a)}
 \frac{\partial Z}{\partial S^b},\\
\frac{\partial W}{\partial S^b}&=
\frac{W^3[(B^av_a)B_b+Zv_b]}{Z(Z+B^aB_a)}\nonumber\\
&\quad{}+\frac{W[Z(1-W^2)-W^2(B^a v_a)^2]}
  {Z(Z+B^aB_a)}\frac{\partial Z}{\partial S^b},\\
\frac{\partial \rho}{\partial S^b}&=-\frac{D}{W^2}
  \frac{\partial W}{\partial S^b},\\
\frac{\partial p}{\partial S^b}&=
  \frac{(B^aB_a)v_b-(B^av_a)B_b}{Z+B^aB_a}
  +\frac{Z+b^2}{Z+B^a B_a} \frac{\partial Z}{\partial S^b},\\
\frac{\partial \epsilon}{\partial S^b}&= -\frac{W}{D}v_b-\frac{p}{D}
  \frac{\partial W}{\partial S^b},\\
\frac{\partial Z}{\partial S^b} &=
 \bigl\{ -[\kappa(Z+B^aB_a) +\rho(B^a B_a+Zc_s^2)]v_b\notag\\
  &\quad{}
  +\rho(B^av_a)(1-c_s^2)B_b\bigr\}
  \big/\mathcal{D},\\
\frac{\partial v^a}{\partial B^b}&=
\frac{Z\gamma^a{}_b(B^av_a)+B^a S_b-2Zv^aB_b}{Z(Z+B^a B_a)}\notag\\
 &\quad{}-\frac{B^a(B^b S_b)+Z^2 v^a}{Z^2 (Z+B^a B_a)}
 \frac{\partial Z}{\partial B^b},\\
\frac{\partial W}{\partial B^b}&=
\frac{W^3(B^av_a)(S_b+Zv_b)-2ZW(W^2-1)B_b}{Z(Z+B^aB_a)}\nonumber\\
&\quad{}+\frac{W[Z(1-W^2)-W^2(B^a v_a)^2]}
  {Z(Z+B^aB_a)}\frac{\partial Z}{\partial B^b},\\
\frac{\partial \rho}{\partial B^b}&=-\frac{D}{W^2}
  \frac{\partial W}{\partial B^b},\\
\frac{\partial p}{\partial B^b}&=
  \frac{[(B^aB_a)+Z(2W^2-1)+W^2(B^av_a)^2]B_b}{W^2(Z+B^aB_a)}\notag\\[5pt]
  &\quad{}-\frac{Z(B^av_a)v_b}{Z+B^aB_a}
  +\frac{Z+b^2}{Z+B^a B_a} \frac{\partial Z}{\partial B^b},\\
\frac{\partial \epsilon}{\partial B^b}&= -\frac{B_b+W^2(B^a v_a)v_b}{WD}
  -\frac{p}{D} \frac{\partial W}{\partial B^b},\\
\frac{\partial Z}{\partial B^b} &=
 -\bigl\{ [\rho Z(2W^2-1-2(W^2-1)c_s^2)\notag\\
  &\qquad\quad{}+\kappa(Z+B^aB_a)+\rho(B^aB_a)]B_b\notag\\
  &\quad{}+W^2(B^av_a)[\kappa(Z+B^aB_a) +\rho(B^a B_a+Zc_s^2)]v_b\notag\\
 &\quad{}-\rho W^2(B^av_a)(1-c_s^2)S_b
  \bigr\} \big/W^2\mathcal{D},\\
\frac{\partial B^a}{\partial B^b}&=\gamma^a{}_b.
\end{align}}

\subsection{Right Eigenvectors in Conserved Variables}
We use Eq.\ \eqref{I-eq:transeigen} of Paper I
with the matrix \eqref{eq:transtocons}
and the eigenvectors in \S\ref{sec:rhoeps} to express the right eigenvectors
in terms of the conserved variables. (Once again: these quantities are
not the eigenvectors for the conservative evolution equations.)
Using $\Rbf$ to denote the right
eigenvectors in conserved variables, we find
\begin{widetext}
\begin{equation}
\begin{gathered}
\Rbf_{\text{entropy}}=\begin{bmatrix}
hW(\kappa-\rho c_s^2)v^a\\
0\\
\kappa\\
\kappa(hW-1)-\rho hW c_s^2
\end{bmatrix},
\quad
\Rbf_{\text{Alf}}=\begin{bmatrix}
-2\sqrt{\rho h^*} B_{21}(B_n+r_1 v_n W^2)\\
-\sqrt{\rho h^*}[B_n B_{32}+B_1 B_{21}+r_1W^2(B_2+v_1B_{21}+v_nB_{32})]\\[2pt]
\sqrt{\rho h^*}[B_n B_{31}-B_2 B_{21}+r_1W^2(B_1-v_2B_{21}+v_nB_{31})]\\[2pt]
0\\
\sqrt{\rho h^*}B_2+v_2r_4\\[2pt]
-\sqrt{\rho h^*}B_1-v_1r_4\\[2pt]
-\rho W B_{21}\\
-B_{21}W(2W\sqrt{\rho h^*}r_1-\rho)
\end{bmatrix},\\[4pt]
\Rbf_\text{constraint}=\begin{bmatrix}
-B^bv_b s^a+(2B_n -v_n B^b v_b)v^a-v_n B^a/W^2\\
s^a-v_n v^a\\
0\\
B_n(2-1/W^2)-2 v_n B^a v_a
\end{bmatrix},
\quad
\Rbf_{\text{mag}}=\begin{bmatrix}
m_\text{1s}s^a + m_\text{1v}v^a + m_\text{1B}B^a\\
\rho h a [B^a(1-y_\text{mag} v_n)-B_n(s^a-y_\text{mag} v^a)]\\[2pt]
-\rho BGB_n/a-\rho^2ah(ya-GW)\\
m_4
\end{bmatrix}.
\end{gathered}
\label{eq:rightanilecons}
\end{equation}
Here $r_1$ and $r_4$ are defined in Eq.\ \eqref{eq:r1}, while
\begin{equation}
B_{21}= B_2 v_1 - B_1 v_2,\qquad
B_{31}= B_n v_1 - B_1 v_n,\qquad
B_{32}= B_n v_2 - B_2 v_n,
\end{equation}
and
\begin{equation}
\begin{split}
m_\text{1s}&=\rho haW(B B^a v_a-\rho h^*a),\\
m_\text{1v}&=\rho h\{B_n B^av_a(ya+2GW)-2aB_n^2-aW[ya(B^aB_a/W^2+\rho h)
+(1-1/c_s^2)\mathcal{G}W]\},\\
m_\text{1B}&=\rho h[B(ya-GW)+2B_n(a^2+G)]/W,\\
m_4&=(\rho/a)\{a^2B^2h-2B_n^2h(a^2+G)+B^2W(2yah-G)+B_nB(G+2hWG-2yah)\\
&\quad+\mathcal{G}Wa^2[hW(1-c_s^2)-1]/c_s^2+\rho ha^3[y-2yh^*W+a(W-h^*)]\}.
\end{split}
\end{equation}
\end{widetext}

While these expressions are somewhat complicated, they are much simpler than
expected based on previous attempts. See, for example, Ref.~\cite{anton2010}
and the discussion in Appendix \ref{app:anton}.

\subsection{Left Eigenvectors in Conserved Variables}

Similarly, applying the transformation matrix defined by the elements in
\S\ref{sec:transmatleft} to the left Anile eigenvectors of
\S\ref{sec:lefteulerian}, we find
\begin{widetext}
\begin{equation}
\begin{gathered}
\Lbf_\text{entropy}=\frac{1}{\rho h c_s^2}\begin{bmatrix}
Wv_b & \gamma_{ba} b^a & h-W & -W
\end{bmatrix},\qquad
\Lbf_\text{constraint}=\frac{1}{1-v_n^2}\begin{bmatrix}
0 & s_b & 0 & 0
\end{bmatrix},\\
\Lbf_\text{Alf} =\dfrac{1}{\sqrt{\rho h^*}}
\left[\begin{matrix}
B_{21}y_\text{Alf}\\
B_2+B_{32}y_\text{Alf}\\
-B_1-B_{31}y_\text{Alf}\\
-B_{21}y_\text{Alf}\sqrt{\rho h^*}\\
-(B_2+B_{32}y_\text{Alf})\sqrt{\rho h^*}\\
(B_1+B_{31}y_\text{Alf})\sqrt{\rho h^*}\\
-B_{21}\\
-B_{21}
\end{matrix}\right]^T,\qquad
\Lbf_\text{mag}=\left[\begin{matrix}
a s_b - BGWB_b/Za  + f_\text{1v}v_b\\
B s_b+g_\text{1B}B_b+g_\text{1v}v_b\\
h_1+\mathcal{G}(\kappa-\rho c_s^2)/\rho^2 c_s^2\\
h_1
\end{matrix}
\right]^T,
\end{gathered}
\label{eq:left_anile_cons}
\end{equation}
\end{widetext}
where
\begin{allowdisplaybreaks}
\begin{align}
f_\text{1v}&=W(-G+BGB_nW/Za^2+\mathcal{G}W^2(\kappa+\rho)/Z\rho c_s^2),\\
g_\text{1B}&=\mathcal{G}\kappa W/\rho Z c_s^2-(a^2+G)/W,\\
g_\text{1v}&=B^av_a W^2g_\text{1B}+W(aB+\mathcal{G}BW^2/Za) ,\\
h_1&=-(f_\text{1v}+ya).
\end{align}
\end{allowdisplaybreaks}
Note that entries 4 -- 6 in $\Lbf_\text{Alf}$ are just $-\sqrt{\rho h^*}$
times entries 1 -- 3.
Note also that finding the relatively simple expressions for $\Lbf_\text{mag}$
required considerable human ingenuity, showing
that sometimes such ingenuity can surpass brute force computer algebra,
at least for now.

\section{Conservative Divergence Cleaning Formulation}
\label{sec:divclean}

The divergence cleaning formulation of MHD is typically presented
as a set of equations in flux-conservative form. To apply our methods, however,
we need to start with an equivalent representation in the comoving formulation.
Fortunately, Ref.~\cite{hilditch2019} has done this for us. The derivation
below is patterned after that work, which shows that these equations are
linear combinations of the flux-conservative ones. Recall, as mentioned
in Paper I, that the eigenvectors of Eq.\ \eqref{I-eq:eigenvalue}
and the determinant in Eq.\ \eqref{I-eq:det_eqns} of Paper I
are invariant under forming linear combinations of the rows of the matrix
$A^a q_a$.

In the divergence cleaning formulation, the Maxwell equation
\eqref{eq:maxwell}
is modified to become
\begin{equation}
\nabla_a(u^b b^a-u^a b^b -g^{ab}\phi)=-\kappa_\tau n^b\phi.
\label{eq:divclean}
\end{equation}
The constant $\kappa_\tau$ is the inverse of the timescale for
driving the divergence constraint exponentially toward zero. Projecting
this equation along $u_b$ now gives the evolution equation for $\phi$
instead of the divergence constraint \eqref{eq:cons3}:
\begin{equation}
u^a \nabla_a\phi+\kappa_\tau W \phi+h^{ab}\nabla_a b_b=0.
\label{eq:evolvephi}
\end{equation}

Projecting Eq.\ \eqref{eq:divclean} perpendicular to $u_b$
gives the replacement for Eq.\ \eqref{eq:maxwell2}:
\begin{equation}
-h_{cb}(b^a\nabla_a u^b-u^a\nabla_a b^b)+b_c \nabla_a u^a+h_{cb}\nabla^b\phi
-\kappa_\tau h_{cb}n^b\phi=0.
\label{eq:maxwell3}
\end{equation}

With the divergence cleaning prescription added to the Maxwell
equation, we cannot derive the energy equation corresponding
to Eq.\ \eqref{eq:pressure} simply by using the First Law \eqref{eq:ds}
as we did in the non-conservative Anile formulation. Instead, we have to go
back to first principles and use the conservation equation
\begin{equation}
u_a\nabla_b T^{ab}=0.
\end{equation}
Subtracting the modified Maxwell equation \eqref{eq:divclean}
projected along $b_b$ from this equation and simplifying, we find
\begin{multline}
\frac{1}{\chi}u^a\left\{\left(h-\frac{p}{\rho}\right)\nabla_a p
-\left[\kappa\left(h-\frac{p}{\rho}\right)-\chi\rho\right]\nabla_a\epsilon
\right\}\\
+\rho h \nabla_a u^a
-b^a(\nabla_a\phi-\kappa_\tau n_a\phi)=0.
\label{eq:energy_intermed}
\end{multline}
Now rewrite Eq.\ \eqref{eq:rhocons} in terms of $(p,\epsilon)$ using
Eq.\ \eqref{eq:chi}:
\begin{equation}
\frac{1}{\chi}u^a(\nabla_a p-\kappa\nabla_a\epsilon)+\rho \nabla_au^a=0.
\label{eq:pepsilon}
\end{equation}
Use this equation to eliminate $u^a\nabla_a\epsilon$ from
Eq.\ \eqref{eq:energy_intermed}. Simplify and find
\begin{equation}
u^a\nabla_a p + \rho h c_s^2\,\nabla_a u^a-\frac{\kappa}{\rho}
b^a(\nabla_a \phi-\kappa_\tau n_a\phi)=0.
\label{eq:pressure_divclean}
\end{equation}

Similarly, using Eq.\ \eqref{eq:pressure_divclean} to eliminate
$u^a\nabla_a p$ from Eq.\ \eqref{eq:pepsilon} and simplifying gives
\begin{equation}
u^a\nabla_a\epsilon+\frac{1}{\rho}(p \nabla_a u^a-b^a\nabla_a\phi
+\kappa_\tau \phi b^a n_a)=0.
\label{eq:epsilon_divclean}
\end{equation}

Finally, we get the Euler equation from the projection of
$\nabla_b T^{ab}=0$ orthogonal to $u^a$. This gives Eq.\ \eqref{eq:euler1}
as before. Now, however, we cannot eliminate the problematic term
$\nabla_b b^b$ using the divergence constraint as we did in
Eq.\ \eqref{eq:combo}. This equation is no longer valid in the
presence of the scalar field $\phi$, being replaced by Eq.\ \eqref{eq:divclean}.
So instead we write
\begin{equation}
\nabla_a b^a=(h^{ab}-u^a u^b)\nabla_a b_b=h^{ab}\nabla_a b_b
+u^a b^b \nabla_a u_b.
\label{eq:divbalt}
\end{equation}
It is convenient not to introduce another term containing $u^a \nabla_a u_b$,
so consider
\begin{align}
0&=b^a\nabla_b T_a{}^b\notag\\
&=b^a\nabla_a p+(\rho h + b^2)b^a u^b\nabla_b u_a-b^2\nabla_a b^a
  \notag\\
&=b^a\nabla_a p+\rho h \nabla_a b^a -(\rho h+b^2) h^{ab}\nabla_b b_a,
\label{eq:gradu_to_p}
\end{align}
where we have used Eq.\ \eqref{eq:divbalt} to get the last equation.
We now eliminate $\nabla_a b^a$ from Eq.\ \eqref{eq:euler1} using this relation
and find
\begin{multline}
\rho h^* h^d{}_b u^a\nabla_a u^b + h^{da}\nabla_a p\\
 + b^a\left[h^d{}_b
(\nabla^b b_a-\nabla_a b^b) +\frac{b^d}{\rho h}\nabla_a p\right]
-\frac{h^*}{h}b^dh^{ab}\nabla_b b_a =0.
\label{eq:euler2mod}
\end{multline} 
This equation is identical to Eq.\ \eqref{eq:euler2} except for the last
term, which is proportional to the divergence constraint.

To enforce the condition $u^a u_a=-1$, in the above equations
we again use the identity \eqref{eq:div}.
Then Eqs.\ \eqref{eq:euler2mod}, \eqref{eq:maxwell3},
\eqref{eq:pressure_divclean}, \eqref{eq:epsilon_divclean},
and \eqref{eq:evolvephi}
can be written as a quasi-linear system (Eq.~\ref{I-eq:system} of Paper I)
for the variables $(u^a,b^a,p,\epsilon,\phi)$,
where the matrices $A^a$ are
\begin{widetext}
\begin{equation}
A^a=
\begin{bmatrix}
u^a h_c{}^b & (b^b h_c{}^a -b^a h_c{}^b)/(\rho h^*)-b_c h^{ba}/(\rho h) &
 h_c{}^a/(\rho h^*)+b_c b^a/(\rho^2 h h^*) & 0 &0\\[3pt]
-b^a h_c{}^b +b_c h^{ba} & h_c{}^b u^a & 0 & 0 & h_c{}^a\\[3pt]
\rho h c_s^2 h^{ba} & 0    & u^a    & 0 & -\kappa b^a/\rho\\[3pt]
(p/\rho)h^{ba} & 0 & 0   & u^a & - b^a/\rho\\[3pt]
0 & h^{ba} & 0 & 0 & u^a
\end{bmatrix}.
\label{eq:amatrix_divclean}
\end{equation}
\end{widetext}

\subsection{Characteristic speeds}
The characteristic speeds are
the zeros of the determinant
of the characteristic matrix $A^a q_a$.
The determinant is evaluated in Appendix \ref{app:divclean} and we find up to
a constant factor
\begin{multline}
\det(A^a q_a)=a G(B^2-\rho h^* a^2)\\
  \times\{a^2G b^2-\rho h a^4+c_s^2[\rho h a^2(a^2+G)- B^2 G]\}.
\label{eq:detdiv}
\end{multline}
Comparing with Eq.\ \eqref{eq:detfinal},
we see that the Alfv\'en and magnetosonic speeds are the same as in
the non-conservative formulation. However, the entropy eigenvalue ($a=0$) is no
longer degenerate with the constraint eigenvalue. The constraint
eigenvalue is replaced by two
eigenvalues, reflecting the increase from an 8- to a 9-dimensional
system. We call these the scalar eigenvalues since they arise
from introducing the scalar field. They
are the roots of $G=q^a q_a=1-y^2=0$.
So $y_\text{scalar}=\pm 1$, which are the wave speeds
for any choice of unit timelike and normal
vectors to define $q^a$.

\subsection{Comoving Eigensystem}

\subsubsection{Right Eigenvectors in the Comoving Frame}

The right eigenvectors follow from solving
$(A^a q_a)X=0$. Take $X$ to be of the form \eqref{eq:Xvec} with
an extra component $X_9$ at the end. Then the analog of
Eqs.\ \eqref{eq:eigveceqns2} is
\begin{subequations}
\label{eq:eigcleaneqns}
\begin{align}
a X_c+\frac{1}{\rho h^*}\Big(b^b Y_b \,h_{ac}q^a -B Y_c \quad &\notag\\
+q^a h_{ac}X_7+\frac{B b_c}{\rho h} X_7
-\frac{h^* b_c}{h}q^bY_b
\Big)&=0,\label{subcleanaa}\\
B X_c-b_c \,q^aX_a- a Y_c -h_{ac}q^aX_9&=0,\label{subcleanbb}\\
\rho h c_s^2 q^a X_a +a X_7-\frac{\kappa B}{\rho}X_9&=0,\label{subcleancc}\\[2pt]
\frac{p}{\rho}q^a X_a +a X_8-\frac{B}{\rho}X_9&=0,\label{subcleandd}\\[2pt]
q^aY_a+aX_9&=0.\label{subcleanee}
\end{align}
\end{subequations}
These equations differ from \eqref{eq:eigveceqns2} in the extra last term
in the first four equations as well as the new fifth equation.

\paragraph{Entropy, Alfv\'en and Magnetosonic Eigenvectors}
The (non-degenerate) entropy eigenvector corresponding to $a=0$,
as well as the Alfv\'en and magnetosonic vectors, are the same as
the expressions given in \S\ref{sec:comoving} with a zero appended.
This follows since Eq.\ \eqref{subcleanee} has a solution $X_9=q^aY_a=0$.
Then Eqs.\ \eqref{eq:eigcleaneqns} reduce to Eqs.\ \eqref{eq:eigveceqns2}
and have the same solutions found there,
since those solutions satisfy $q^aY_a=0$.

\paragraph{Scalar Eigenvectors}
It remains to find the two scalar eigenvectors corresponding to
the eigenvalues given by $G=1-y^2=0$. Since the equations are
linearly dependent when the eigenvalue is inserted, it is easy
to get $0=0$. One way of proceeding is as follows:
\begin{itemize}
\item
Use Eq.\ \eqref{subcleanee} to eliminate $X_9$ from the other equations.
\item
Use Eq.\ \eqref{subcleancc} to eliminate $X_7$ from Eq.\ \eqref{subcleanaa}.
\item
Dot Eq.\ \eqref{subcleanaa} with $b^c$ and solve the resulting equation
for $q^c Y_c$.
\item
Dot Eq.\ \eqref{subcleanaa} with $q^c$, use the expression for $q^c Y_c$
from the previous step, and solve for $b^c Y_c$.
\item
Also use the expression for $q^c Y_c$ in Eq.\ \eqref{subcleanbb}
and solve the equation for $Y_c$ in terms of $X_c$, $q^c X_c$, and
$b^c X_c$.
\item
In Eq.\ \eqref{subcleanaa}, make the substitutions for $q^c Y_c$,
$b^c Y_c$, and $Y_c$. Dot with $b^c$ and solve for $b^c X_c$. Substitute
this back in the equation and solve for $X_c$. An overall factor of
$q^c X_c$ is an arbitrary normalization. Choose it to remove the
denominator of $X_c$ to find
\begin{equation}
X_c=\kappa_\rho B h_{cb}q^b+\rho a^2(1-c_s^2) b_c,
\end{equation}
where
\begin{equation}
\kappa_\rho\equiv\kappa+\rho c_s^2.
\end{equation}
\item
Use this expression to back-solve for the other quantities.
\end{itemize}
The final expression for the eigenvector is
\begin{multline}
X_\text{scalar}=\\
\begin{bmatrix}
\kappa_\rho B h_{ab}q^b+\rho a^2(1-c_s^2) b_a\\[2pt]
-\kappa_\rho a B b_a+h_{ab}q^b[\kappa_\rho B^2/a+\rho^2 h a(1-c_s^2)]\\[2pt]
-\kappa_\rho a B \rho h\\[2pt]
-a B [p+\rho(h-\chi)]\\[2pt]
-\rho^2 h a^2(1-c_s^2)
\end{bmatrix}.
\end{multline}
This agrees with Eq.\ (36) in \cite{hilditch2019} when expressed in a spatial
basis in the comoving frame.

\subsubsection{Left Eigenvectors in the Comoving Frame}
\label{sec:left_comov_clean}

The equations $L(A^a q_a)=0$ for the left eigenvectors are very similar to
the case of the Anile formulation treated in \S\ref{sec:leftcomoving}.
There is an additional scalar $L_9$ in the equation corresponding to
\eqref{eq:Lvec}. The equations are the same as those in
Eq.\ \eqref{eq:leigveceqns2} except that there is an additional term
in Eq.\ \eqref{subleqbb} of the form
\begin{equation}
h^* h_{ca}q^a(L^bb_b/h-\rho L_9)
\end{equation}
and there is a new fifth equation
\begin{equation}
B(\kappa L_7+L_8)-\rho a L_9-\rho Y^bq_b=0.
\end{equation}

\paragraph{Entropy Eigenvector}
The entropy eigenvector corresponding to $a=0$ is again the only one
with $L_8\neq 0$. Equation \eqref{subleqcc} again implies $L_b=0$, and
solving the remaining equations gives
\begin{equation}
L_\text{entropy}=
\begin{bmatrix}
0 & h_{ab}q^b \chi B/G & -p/\rho & \rho h c_s^2 & 0
\end{bmatrix}.
\end{equation}

\paragraph{Scalar Eigenvectors}
For the other eigenvectors, we set $L_8=0$ and use the remaining two
scalar equations to eliminate $L_7$ and $L_9$ from the two vector equations.
For the scalar eigenvector ($G=0$), the Anile case \eqref{eq:ldegen}
suggests the ansatz $L_b=0$. We then find
\begin{equation}
L_\text{scalar}=\begin{bmatrix}
0 & h_{ab}q^b & 0 & 0 & -a
\end{bmatrix}.
\end{equation}

\paragraph{Alfv\'en Eigenvectors}
Similarly, for
the Alfv\'en eigenvector we try the ansatz that $L_b$ and $Y_b$ are both
orthogonal to $q^a$ and $b^a$. The we find that for $B=-\sqrt{\rho h^*}a$
the eigenvector is the same as
Eq.\ \eqref{eq:lalf} with a zero appended:
\begin{equation}
L_\text{Alf}=\begin{bmatrix}
\sqrt{\rho h^*}\epsilon_{abcd}u^b b^c q^d &
-\epsilon_{abcd}u^b b^c q^d & 0 & 0 & 0
\end{bmatrix}.
\label{eq:lalf_clean}
\end{equation}
The other eigenvector has no minus sign and the appropriate
change in $q^d$.
There is an alternative representation in the form of Eq.\ \eqref{eq:alfleftalt}
with a zero appended.

\paragraph{Magnetosonic Eigenvectors}
For the magnetosonic eigenvectors, since Eq.\ \eqref{subeqlaa} is the
same in the divergence cleaning case, we make the ansatz that the
corresponding quantity $L_b$ is the same too, namely the first element
in Eq.\ \eqref{eq:lmagcomov}. We look for $Y_b$ as a linear combination
of $b_b$ and $h_{ba}q^a$ and find a solution very similar to
Eq.\ \eqref{eq:lmagcomov}. The differences are shown shaded in the following
equation:
{\setlength\delimitershortfall{0pt}  %this gets the bracket high enough for ^T
\begin{equation}
L_\text{mag}=\left[\begin{matrix}
B(a^2+G)b_a/a-\rho h^* a h_{ab}q^b\\[2pt]
(a^2+G)b_a- Bh_{ab}q^b[1 \colorbox{mygray}{${}
  -\kappa_\rho\mathcal{G}_\rho/(\rho G)$}]\\[2pt]
\mathcal{G}_\rho\\[2pt]
0\\
\colorbox{mygray}{$-\mathcal{G}_\rho B[c_s^2/a+\kappa_\rho a/(\rho G)]$}
\end{matrix}
\right]^T,
\label{eq:lmagdivclean}
\end{equation}}%
where
\begin{equation}
\mathcal{G}_\rho=\mathcal{G}/\rho h c_s^2.
\end{equation}

\subsection{Eigenvectors in the Eulerian frame}

\subsubsection{Right Eigenvectors in the Eulerian frame}
\label{sec:rightcleaneuler}

As for the non-conservative formulation,
the transformation from the comoving to the Eulerian right eigenvectors
is given by the Jacobian matrix $\partial(v^a,B^a)/\partial(u^b,b^b)$.
This matrix acts on the vector part of each eigenvector. The scalar
part of the eigenvectors is unchanged.
Since all the comoving
right eigenvectors in the divergence cleaning formulation
are the same as for the non-conservative formulation (with a zero appended)
except $X_{\text{scalar}}$,
the eigenvectors in the Eulerian frame are
the same as Eqs.\ \eqref{eq:xfontent}, \eqref{eq:alffont}, and
\eqref{eq:xmagfinal} with a zero appended. The scalar eigenvector is
the only one that is different. There are two eigenvectors corresponding
to the roots $y=\pm 1$ of $G=1-y^2=0$:
\begin{multline}
\xbf_{\text{scalar}}=\\
\begin{bmatrix}
[\kappa_\rho BWF^a+(1-c_s^2)\rho a^2 B^a]/W^2\\[2pt]
n^bq_b\kappa_\rho B B^a/W+[\kappa_\rho B B_n+(1-c_s^2)\rho^2 h a^2W]
  F^a/a\\[2pt]
-\kappa_\rho \rho h B a\\[2pt]
-[p+\rho(h-\chi)]Ba\\[2pt]
-\rho^2 h a^2(1-c_s^2)
\end{bmatrix},
\label{eq:divconstraint}
\end{multline}
where $F^a=\gamma^{ab}q_b+(n^bq_b) v^a$. Note that $F^a=s^a-yv^a$ and $n^bq_b
=-y$ with $y=\pm 1$, but we will not make these substitutions yet.

Applying the transformation \eqref{eq:inversekappa} to
Eq.\ \eqref{eq:divconstraint} to change the
thermodynamic variables from $(p,\epsilon)$ to $(\rho,\epsilon)$, we
find that only the third entry changes:
\begin{equation}
-\kappa_\rho \rho h B a\to -(\kappa+\rho)\rho B a.
\end{equation}

Note that Ref.\ \cite{hilditch2019} was not able to find a simple
expression for $\xbf_{\text{scalar}}$ using their dual frames approach.

\subsubsection{Left Eigenvectors in the Eulerian frame}
\label{sec:leftdiveulerian}

The transformation \eqref{eq:a0} for the divergence cleaning case has some
additional terms. First, expand \eqref{eq:a0} to a $5\times5$ matrix by padding
it with zeros. Then add the following terms:
\begin{equation}
\begin{bmatrix}
0 & Wb^av^b/(\rho h) & 0 & 0 & 0\\[2pt]
0 & 0 & 0 & 0 &-WH^a\\[2pt]
0 & 0 & 0 & 0 & -\kappa (b^a n_a)/\rho\\[2pt]
0 & 0 & 0 & 0 & -(b^a n_a)/\rho\\[2pt]
0 & -Wv^b & 0 & 0& -W
\end{bmatrix}
\end{equation}
Applying this transformation to the left eigenvectors in
\S\ref{sec:left_comov_clean} and changing
from $(p,\epsilon)$ to $(\rho,\epsilon)$, we find
\begin{align}
&\lbf_\text{entropy}=\notag\\
&\begin{bmatrix}
0 & B \gamma_{ab}q^b/G & -pW/\rho & \rho h W c_s^2
  & b^a n_a-Bq^a n_a/G
\end{bmatrix},\\
&\lbf_\text{scalar}=
\begin{bmatrix}
0 & \gamma_{ab}q^b & 0 & 0 & -q^a n_a
\end{bmatrix}.
\end{align}

Since the comoving Alfv\'en eigenvector \eqref{eq:lalf_clean} is essentially
the same as Eq.\ \eqref{eq:lalf}, we find that the transformation
to the Eulerian frame gives the same form as Eq.\ \eqref{eq:lalffinal} for
the first 8 components.
The change to $(\rho,\epsilon)$
transforms the seventh and eighth components
to Eq.\ \eqref{eq:lalf_rhoeps}.
Only the ninth (last) component is new,
and it takes the simple form 
\begin{equation}
-(B_2 v_1-B_1 v_2).
\label{eq:left_div_alf_extra}
\end{equation}

The magnetosonic eigenvectors turn out to be the same as
Eq.\ \eqref{eq:lmagfinal} except for some small changes shown shaded below.
\begin{multline}
\lbf_\text{mag} =\\
\left[\begin{matrix}
a(B^aB_a+\rho h W^2)s_a-(S v_a+TB_a)/a\\[3pt]
(aB_n+KB^bv_b)v_a+\dfrac{KB_a}{W^2}+\left(\dfrac{B_n}{W} \colorbox{mygray}{${}
 -\dfrac{\mathcal{G}_\rho \kappa_\rho B}{G\rho}$} \right)s_a\\[8pt]
K+\dfrac{GBB_n}{\rho h a^2}\\[8pt]
0\\
\colorbox{mygray}{$
 \dfrac{B K(G\rho-\mathcal{G}_\rho \kappa_\rho)+G B_n[\rho(a^2+G)-
 \mathcal{G}_\rho \kappa]}{G\rho a}$}
\end{matrix}
\right]^T.
\label{eq:lmag_clean}
\end{multline}
Transforming from $(p,\epsilon)$ to $(\rho,\epsilon)$
changes the third and fourth terms to those shown
in Eq.\ \eqref{eq:lmag_last_two}.

\subsection{Eigenvectors for Conserved Variables}

\subsubsection{Right Eigenvectors for Conserved Variables}
As mentioned already in \S\ref{sec:rightcleaneuler}, with the exception
of the scalar eigenvectors, the right eigenvectors
are the same as for the Anile case with a zero appended.
Since the
transformation matrix \eqref{eq:transtocons} is the same, the eigenvectors
for the conserved variables are the same as
in Eq.~\eqref{eq:rightanilecons}, with the exception of
$\Rbf_\text{constraint}$, which gets replaced by the scalar eigenvectors.

The scalar eigenvectors transform to
\begin{widetext}
\begin{equation}
\Rbf_\text{scalar}=\begin{bmatrix}
-\frac{1}{aW}[(y\kappa_B +2\kappa_\rho aBB_n)B^a +W^2\kappa_{Bv}(s^a+yv^a)
  -2W\kappa_\rho B B_n^2v^a]\\[3pt]
\kappa_\rho y B B^a/W+(s^a-yv^a)[\kappa_\rho B B_n +(1-c_s^2)\rho^2 a^2
  hW]/a\\[3pt]
\kappa_\rho y\rho B -(1-c_s^2)\rho^2 a B_n\\[3pt]
\{\kappa_\rho B[2B_n^2+\rho a(ah^*-y)]-\kappa_B B-2\kappa_{Bv}yW
+(1-c_s^2)\rho^2 a^2B_n\}/a\\[3pt]
-(1-c_s^2)\rho^2a^2h
\end{bmatrix},
\end{equation}
\end{widetext}
with $y=\pm 1$. Here
\begin{equation}
\begin{split}
\kappa_B&=\kappa_\rho B^2+(1-c_s^2)\rho^2 a^2 h,\\
\kappa_{Bv}&=\kappa_B B^av_a-\kappa_\rho \rho a B h^*.
\end{split}
\end{equation}

\subsubsection{Left Eigenvectors for Conserved Variables}

For the divergence cleaning case, the field $\phi$ is both a primitive and
a conservative variable. Thus, the inverse transformation matrix
derived in \S\ref{sec:transmatleft} simply gets an extra row and column
of zeros, with a 1 in the lower-right corner.
Applying the transformation to the left eigenvectors given in
\S\ref{sec:leftdiveulerian}, we get
{\setlength\delimitershortfall{0pt}  %this gets the bracket high enough for ^T
\begin{align}
\Lbf_\text{entropy}&=\frac{1}{\rho h c_s^2}
\left[\begin{matrix}
Wv_b \\ \gamma_{ba} b^a -(B_n/GW)s_b \\ h-W \\ -W \\ W B^a v_a-B_nv_n/GW
\end{matrix}\right]^T,\\
\Lbf_\text{scalar}&=\frac{1}{1-v_n^2}
\left[\begin{matrix}
0 \\ s_b \\ 0 \\ 0 \\ y_\text{scalar}
\end{matrix}\right]^T.
\end{align}}%
Since $\lbf_\text{Alf}$ is the same as for the Anile case except for
the additional term \eqref{eq:left_div_alf_extra}, the eigenvector
$\Lbf_\text{Alf}$
for the conserved variables is the same as in Eq.~\eqref{eq:left_anile_cons}
with the additional element $-B_{21}\sqrt{\rho h^*}$ appended.

The magnetosonic eigenvectors are almost identical to those in
Eq.~\eqref{eq:left_anile_cons}. The only differences are exactly
the two additional terms shown shaded in Eq.\ \eqref{eq:lmag_clean}.
The first additional term gets added to the second element in
Eq.~\eqref{eq:left_anile_cons}, while the second additional term becomes
the new fifth element.

\subsection{Eigenvectors for the Valencia Formulation}

It is perhaps surprising that the eigenvalues
and eigenvectors for the divergence
cleaning formulation are so similar to those of the Anile formulation.
Note that
the Valencia formulation does not follow from setting the
scalar field $\phi$ to zero in the divergence cleaning system. Nevertheless,
the eigenvalues again are almost all
identical to those in Eq.~\eqref{eq:detfinal}. However, Ref.~\cite{schoepe2018}
shows that even in the comoving frame, the eigenvectors differ
significantly from those given in \S\ref{sec:comovingdecomp}.
The expected symptoms of weak hyperbolicity, such as lack of convergence,
have not been reported in simulations using the Valencia formulation.
This could be because of the difficulties in showing strict convergence
in simulations containing magnetic field instabilities or shocks.
Many numerical implementations are able to use constrained
transport~\cite{evans1988,balsara1999,toth2000},
which guarantees conservation of $\nabla\cdot\Bbf$ to machine precision.
If the constraint were identically satisfied, this could
explain why the system is actually strongly hyperbolic in the constraint-satisfying
subspace. Then simulations would have to be at machine precision to see
signs of weak hyperbolicity. This speculation is obviously difficult
to demonstrate in practice.

Since the Valencia formulation is weakly hyperbolic and not all
simulation schemes allow constrained transport, including relativistic
discontinuous Galerkin schemes,
we have not worked out its decomposition here. However, the methods
of this paper could be used to find the decomposition in terms
of the conserved variables used in simulations.

\section{Conclusions and Future Work}
\label{sec:conclusions}

In this paper, we have presented a straightforward method to derive
the characteristic decomposition for GRMHD. The resulting eigenvectors
are simple enough to be used in numerical simulations. This simplicity
is somewhat surprising given the complexity of previous attempts.

The key innovation that makes these results possible is the
introduction of a quasi-invertible transformation, as described
in \S\ref{sec:quasiinvert}. This transformation allows us
to transform between the variables $\{u^a,b^a\}$ in the comoving frame
and the variables $\{v^a,B^a\}$ in the Eulerian frame in such away that
the correct eigenvectors are obtained.
Together with the new transformation law for left eigenvectors
\eqref{eq:ltrans}, the quasi-invertible transformation allows
the solution of the problem.

We have first derived the decomposition in terms of conserved variables
for the non-conserved formulation of Anile. This decomposition is not intended
to be used in simulations because this formulation is not equivalent
to the flux-conservative formulation used in practice~\cite{schoepe2018}.
We give this decomposition solely to make contact with previous work,
and because the results are very close to the formulation that
uses a divergence-cleaning scalar field with the flux-conservative
form.
 
We are currently implementing this decomposition in the
\texttt{SpECTRE} code~\cite{spectrecode} and will test its
performance with a full-wave Riemann solver such as the Marquina
scheme~\cite{donat1996,aloy1999}.

An important future step is
to study the impact of degeneracies on the decomposition.
Degeneracies have been carefully studied and circumvented
in the comoving~\cite{anile1989} formulation and also in the
work of Ref.~\cite{anton2010}. However, such a treatment will have to
be redone for the conservative case treated here and will
be the subject of a future paper. Degeneracies
only occur for very specific orientations of the magnetic field
and so hopefully at only a small number of grid points in
a realistic simulation.
In this case, as a temporary measure we can get the eigenvectors numerically
from the characteristic matrix itself.

\begin{acknowledgments}
We have greatly benefited from the Mathematica notebooks that
accompany Refs.~\cite{schoepe2018} and \cite{hilditch2019}.
While the details of our calculations are very different,
the Mathematica manipulations have many similarities.
These notebooks use the xTensor package~\cite{xtensor}.
This work was supported in part by NSF grants
PHY-2308615 and OAC-2513338 and by NASA award 80NSSC26K0340 at Cornell.
This work was also supported in part by the Sherman
Fairchild Foundation at Caltech and Cornell.
\end{acknowledgments}

\appendix

\section{Comparison with Previous Methods}
\label{app:previous}
\subsection{The Method of Anton et al.}
\label{app:anton}

In Ref.\ \cite{anton2010}, Ant{\'o}n et al.\ found the characteristic
decomposition for MHD in special relativity by a transformation method.
They start with the covariant Anile eigenvectors for the 10 variables
$U_\text{cov}=(u^a,b^a,p,s)$. The particular choice of the two thermodynamic
variables is not relevant for the discussion here. They use the same
set of 8 conserved variables $U_\text{cons}$ that we use,
defined in \S\ref{sec:consvariables}. They derive the eigensystem
along the $x$-axis, so they can eliminate $B^x$ since it satisfies the
trivial evolution equation $\partial B^x/\partial t=0$,
leaving 7 variables. They then compute the $7\times 10$ transformation
matrix $\partial U_\text{cons}/\partial U_\text{cov}$ for the right
eigenvectors by direct differentiation, using  the constraints
\eqref{eq:constraints} and the 0-component of Eq.\ \eqref{eq:maxwell},
which is the divergence constraint.

For the inverse transformation, needed for the left eigenvectors, they
introduce an intermediate set of 7 primitive variables $V=(u^x,u^y,u^z,
b^y,b^z,p,\rho)$. Here, the vector quantities are a subset of the covariant
variables that are easy to work with in special relativity.
The inverse transformation is than essentially
computed by using the matrix product
\begin{equation}
\frac{\partial U_\text{cov}}{\partial V}
 \frac{\partial V}{\partial U_\text{cons}}.
\label{eq:anton}
\end{equation}
For this procedure to work, we require that the $10 \times 7$ matrix
defined by the product \eqref{eq:anton} serve as the matrix inverse
of $\partial U_\text{cons}/\partial U_\text{cov}$. But as we have seen,
a rectangular matrix may only have a one-sided inverse in general.
We have not checked whether the matrices given in \cite{anton2010}
function as inverses, since the calculation seems daunting with the
information given. However, even if the method is consistent,
it does not seem possible to apply it in general relativity.
Here the relation between Eulerian and comoving frames is more
complicated, which is why we introduced the idea of quasi-invertibility. 

More important, as mentioned earlier, the Anile equations do not
correspond~\cite{schoepe2018} to the conservation equations,
which are those typically
used in numerical simulations. Simply transforming
the Anile equations to conserved variables does not lead to the correct
decomposition.

Finally, the transformations in Ref.\ \cite{anton2010} are extremely
complicated, leading to eigenvectors covering several journal pages.
For example, their equations (163) -- (166) correspond to
our short
expression for $\Lbf_\text{entropy}$ in eq.~\eqref{eq:left_anile_cons}.
Their equations contain derivatives of the quantity $Z$, which are evaluated
only for the case of an ideal fluid in the thesis\cite{antonthesis}. We have
evaluated these derivatives for a general equation of state in
\S\ref{sec:transmatleft}. We have substituted our expressions into
their equations (163) -- (166) and managed to simplify the equations.
After many cancellations, two of
their equations reduce to the first and fourth elements of
$\Lbf_\text{entropy}$. We have not tried to track down the origin
of the differences in the other two elements, since
our methods lead to much simpler expressions as it is.
Note that three of the elements in $\Lbf_\text{entropy}$ are the same
as in hydrodynamics---the $\Bbf$ field cancels out.

\subsection{The Method of Schoepe et al.}
\label{app:schoepe}

Schoepe et al.~\cite{schoepe2018} introduced an algorithm for
characteristic decomposition in relativity that relies on a
dual frames approach.
The two frames in the algorithm are again the comoving and the Eulerian frames
of the fluid. The formalism is derived from the
dual foliation formalism~\cite{hilditch2015}, but because the
comoving frame does not correspond to a foliation, the name
was changed to
dual frames. Note that this treatment
is not related to the earlier use of ``dual frames''
in Ref.\ \cite{lindblom2006},
which forms the basis of the SpEC numerical relativity
code~\cite{spec_skip}.

In Refs.~\cite{schoepe2018,hilditch2019},
the dual-frame formalism is used very successfully to address
the hyperbolicity of various formulations of relativistic
fluid equations. In particular, they show the surprising result
that the most commonly used
formulation of GRMHD is only weakly hyperbolic, and so does not have
a well-posed initial value problem. 

For numerical work, however, there are several
drawbacks to the dual-frames approach.
First, the spatial basis in the Eulerian frame is treated in a non-standard
way to simplify the algebra, and one has to translate
those results to the standard Eulerian coordinate basis. Also,
from the point of view of numerical work,
non-standard choices of primitive variables are made.
These are relatively minor drawbacks.
From our perspective, more serious is that
the computations are quite complicated and require
computer algebra essentially throughout, especially for GRMHD.
Thus, extending the dual-frames results from the decomposition 
for the primitive variables to that of the conserved variables
in a useful form for numerical work
is a somewhat daunting task that remains to be carried out.
By contrast, the quasi-invertible transformation technique introduced
here leads to much simpler calculations and allows the complete
determination of the GRMHD decomposition in term of conserved variables,
as shown in this paper.

\section{Evaluating the Characteristic Determinant}
\label{app:a}

\subsection{Non-Conservative Formulation}

In the usual comoving approach in the literature, one typically
deals with a $10\times 10$ characteristic
determinant corresponding to the variables $(u^a,b^a)$ and two thermodynamic
variables. The standard form of the determinant is given in \cite{anile1989},
with the remark
``Now it is easy to show that \dots'' and
the value of the determinant is written down without derivation.
Subsequent authors have also just quoted
the answer, either repeating Anile's remark or without comment.
The determinant was actually first evaluated in 1966
by Bruhat~\cite{bruhat1966}, with a derivation.

In our case, the characteristic matrix has rank 8 since we
have explicitly enforced the two constraints on $u^a$ and $b^a$.
Equation~\eqref{eq:amatrix} gives
\begin{widetext}
\begin{equation}
A^a q_a=
\begin{bmatrix}
u^a q_a h_c{}^b & q_a(b^b h_c{}^a -b^a h_c{}^b)/(\rho h^*) &
 q_a[h_c{}^a/(\rho h^*)+b_c b^a/(\rho^2 h h^*)] & 0\\[3pt]
q_a(-b^a h_c{}^b +b_c h^{ba}) & h_c{}^b u^a q_a & 0 & 0\\[3pt]
\rho h c_s^2 h^{ba}q_a & 0    & u^aq_a    & 0\\[3pt]
(p/\rho)h^{ba}q_a & 0 & 0   & u^aq_a
\end{bmatrix}.
\end{equation}
Formally, this is a $10\times 10$ matrix with rank 8, so we must be careful
in evaluating the determinant. Since the determinant is a scalar, we
can evaluate it in the local comoving frame of the fluid, where $h^b{}_c\to
\delta^i{}_j$, a $3\times 3$ matrix.
Using the definitions \eqref{eq:defns}, we get
\begin{equation}
\det(A^aq_a)=a \det\begin{bmatrix}
a \delta^i{}_j & (-B\delta^i{}_j+b^i q_j)/(\rho h^*) &
(B b_j+\rho h q_j)/(\rho^2 h h^*) \\[3pt]
-B\delta^i{}_j+q^i b_j& a \delta^i{}_j & 0 \\[3pt]
\rho h c_s^2 q^i & 0    & a
\end{bmatrix}.
\label{eq:3by3det}
\end{equation}
\end{widetext}
We evaluate the determinant using the formula for the determinant of
a partitioned matrix:
\begin{equation}
\det
\begin{bmatrix}
P & Q\\
R & S
\end{bmatrix}
=
\det S \det(P-QS^{-1}R).
\label{eq:partition}
\end{equation}
In our case,
\begin{equation}
\begin{split}
P &= a \delta^i{}_j,\\
Q &= \begin{bmatrix}
(-B\delta^i{}_j+b^i q_j)/(\rho h^*) & (B b_j+\rho h q_j)/(\rho^2 h h^*)
\end{bmatrix},\\
R &= \begin{bmatrix}
-B\delta^i{}_k+q^i b_k\\[3pt]
\rho h c_s^2 q^i
\end{bmatrix},\\[3pt]
S&=\begin{bmatrix}
a \delta^k{}_i & 0\\
0    & a
\end{bmatrix}.
\end{split}
\label{eq:pqrs}
\end{equation}
Here we have arranged the tensor indices to facilitate the matrix
multiplications in Eq.\ \eqref{eq:partition}.
We find
\begin{equation}
Q S^{-1}=\begin{bmatrix}
(-B \delta^k{}_j+b^k q_j)/(\rho h^* a) &
(B b_j+\rho h q_j)/(\rho^2 h h^* a)
\end{bmatrix}
\end{equation}
and
\begin{equation}
\begin{gathered}
Q S^{-1} R =
(B^2 \delta^i{}_j + w^i q_j+z^i b_j)/(\rho h^* a),\\
w^i\equiv B b^i-(b^2+\rho h c_s^2)q^i,\quad
z^i\equiv (1-c_s^2)B q^i,
\end{gathered}
\label{eq:qsr}
\end{equation}
and hence
\begin{equation}
\det(A^a q_a) =
a^5\det(T \delta^i{}_j + w^i q_j+z^i b_j)/(\rho h^* a)^3,
\label{eq:intermed-det}
\end{equation}
where
\begin{equation}
T\equiv -(B^2-\rho h^* a^2).
\end{equation}
The remaining determinant in Eq.\ \eqref{eq:intermed-det} is given by the
identity
\begin{multline}
\det\begin{bmatrix}
T\delta^i{}_j+w^i q_j+z^i b_j
\end{bmatrix}\\
=T\left[T^2+T(w^i q_i+z^i b_i)+w^i q_i\, z^j b_j
-w^i b_i\, z^j q_j\right],
\label{eq:appdet}
\end{multline}
which we prove below.
Now we have
\begin{equation}
\begin{split}
w^iq_i&= B^2-q^i q_i(b^2+\rho h c_s^2),\\
z^j b_j &= (1-c_s^2)B^2,\\
w^i b_i &= -\rho h c_s^2 B,\\
z^j q_j &= (1-c_s^2)B q^j q_j.
\end{split}
\end{equation}
In these equations, we can go back from the local rest frame of the fluid
to a general frame by the replacement
\begin{equation}
q^i q_i \to h_{ab}q^a q^b = G+a^2.
\label{eq:qdotq}
\end{equation}
Putting all the pieces together, Eq.\ \eqref{eq:intermed-det} becomes
\begin{multline}
\det(A^a q_a) =
a^2(B^2-\rho h^* a^2)\\
  \times\{a^2G b^2-\rho h a^4+c_s^2[\rho h(a^2+G)- B^2 G]\}/(\rho h^*)^2.
\label{eq:final_det}
\end{multline}
Dropping the constant denominator gives Eq.\ \eqref{eq:detfinal} of the
main text.

To prove the identity \eqref{eq:appdet}, which is a generalization of
the matrix determinant lemma,
note that the determinant is a scalar. The only scalars
that can be formed out of $T$, $w^i$, $z^i$,
$q_i$ and $b_i$ are $T$, $w^i q_i$, $w^i b_i$,
$z^i q_i$ and $z^i b_i$. (Quantities like
$w^i z_i$ cannot appear because the determinant
only multiplies and adds entries together; it cannot raise
or lower indices.) Each term in the determinant consists
of 3 elements multiplied together. These must be 3 $T$'s,
or 2 $T$'s and a pair of vector elements like
$w^i$ and $q_j$,
or 1 $T$ and a set of 4 elements from distinct vectors.
No elements from the same vector can be repeated, since
the determinant uses only one element from each row or
column at a time. This is also the reason that none of
the allowed scalar products can appear squared.
Thus the form of the determinant must be
\begin{multline}
T[a_0 T^2+T(a_1w^i q_i+a_2z^i b_i+a_3
w^i b_i+a_4 z^j q_j)\\
+a_5 w^i q_i\, z^j b_j+a_6 w^i b_i\, z^j q_j].
\end{multline}
The only way that a term like $q_i$ can appear is multiplied
by a $w^j$, since the determinant selects these terms together.
Similarly $b_i$ must appear together with $z^j$.
Thus $a_3=a_4=0$. Now
choose $w^i=(w^1,w^2,0)$, $z^i=(z^1,z^2,0)$,
$b_j=(b_1,b_2,0)$ and $q_j=(q_1,0,0)$ and evaluate
the determinant to determine $a_0,a_1,\ldots$
This gives the result \eqref{eq:appdet}.

\subsection{Divergence Cleaning Formulation}
\label{app:divclean}

Equation~\eqref{eq:amatrix_divclean} gives in the comoving frame
\begin{widetext}
\begin{equation}
\det(A^aq_a)=\det\begin{bmatrix}
a \delta^i{}_j & (-B\delta^i{}_j+b^i q_j)/(\rho h^*)-q^i b_j/(\rho h) &
(B b_j+\rho h q_j)/(\rho^2 h h^*) & 0 & 0\\[3pt]
-B\delta^i{}_j+q^i b_j& a \delta^i{}_j & 0 & 0 & q_j \\[3pt]
\rho h c_s^2 q^i & 0    & a & 0 &  -\kappa B/\rho\\[3pt]
(p/\rho)q^i & 0 & 0 & a & -B/\rho\\[3pt]
0 & q^i & 0 & 0 & a
\end{bmatrix}.
\end{equation}
Multiply the last column by $q^i/a$ and subtract from the second column to
eliminate $q^i$ in the bottom row. This gives
\begin{equation}
\det(A^aq_a)=a^2\det\begin{bmatrix}
a \delta^i{}_j & (-B\delta^i{}_j+b^i q_j)/(\rho h^*)-q^i b_j/(\rho h) &
(B b_j+\rho h q_j)/(\rho^2 h h^*) \\[3pt]
-B\delta^i{}_j+q^i b_j& a \delta^i{}_j-q^i q_j/a & 0\\[3pt]
\rho h c_s^2 q^i & \kappa B q^i/(\rho a)  & a
\end{bmatrix}.
\end{equation}
Following the same procedure as we used to evaluate
Eq.\ \eqref{eq:3by3det}, we see that Eq.\ \eqref{eq:pqrs}
is replaced by 
\begin{equation}
\begin{alignedat}{2}
P &= \phantom{\big[}a \delta^i{}_j\phantom{\big]},\qquad &
Q &= \begin{bmatrix}
(-B\delta^i{}_j+b^i q_j)/(\rho h^*)-q^i b_j/(\rho h) &
(B b_j+\rho h q_j)/(\rho^2 h h^*)
\end{bmatrix},\\[4pt]
R &= \begin{bmatrix}
-B\delta^i{}_k+q^i b_k\\[3pt]
\rho h c_s^2 q^i
\end{bmatrix},\qquad &
S&=\begin{bmatrix}
a \delta^k{}_i -q^k q_i/a& 0\\[3pt]
\kappa B q^k/(\rho a)  & a
\end{bmatrix}.
\end{alignedat}
\end{equation}
\end{widetext}
Using the matrix determinant lemma, we find that
\begin{equation}
\begin{split}
\det(S)&=a\det(a \delta^k{}_i -q^k q_i/a)=a^2 \theta,\\
\theta&\equiv a^2-q_i q^i.
\end{split}
\end{equation}
Using, for example, partitioned matrix techniques again, we can compute
\begin{equation}
S^{-1}=\frac{1}{a \theta}\begin{bmatrix}
\theta\delta^i{}_j+q^i q_j & 0 \\[3pt]
-\kappa B q^i/\rho & \theta
\end{bmatrix}.
\end{equation}
So
\begin{widetext}
\begin{equation}
Q S^{-1}= \left[\begin{matrix}
-\rho h[\kappa B q^k q_j+\rho \theta(B\delta^k{}_j-b^kq_j)+(\kappa B^2+\rho^2h^*
a^2)q^kb_j]/(\rho^3 h h^* \theta a)\\
(Bb_j+\rho h q_j)/(\rho^2 h h^*a)
\end{matrix}
\right]^T.
\end{equation}
\end{widetext}
We then find that $Q S^{-1} R$ is given by the same expression as
Eq.\ \eqref{eq:qsr}, and hence the only difference from
Eq.\ \eqref{eq:intermed-det} is that $\det(S)=a^5$ is replaced by
$a^2\det(S)=a^4 \theta$. Since $\theta=a^2-q_i q^i=-G$ by Eq.\ \eqref{eq:qdotq},
we get Eq.\ \eqref{eq:final_det} but with the initial factor of $a^2$ replaced
by $a G$. This gives Eq.\ \eqref{eq:detdiv} of the main text.

\bibliography{references}
\end{document}